\documentclass[aps,prev,twocolumn,preprintnumbers,floatfix,nofootinbib]{revtex4-1}
\pdfoutput=1
\usepackage{graphicx}
\usepackage{bm}
\usepackage{times}
\usepackage{slashed}
\usepackage{color}
\usepackage{aas_macros}
\usepackage{slashed}
\usepackage{lipsum}
\usepackage{subfigure}
\usepackage{multirow}
\usepackage{amsmath}
\usepackage{array} 
\usepackage{varwidth} 

\newcommand{\be}{\begin{equation}}
\newcommand{\ee}{\end{equation}}
\newcommand{\bea}{\begin{eqnarray}}
\newcommand{\eea}{\end{eqnarray}}

\begin{document}

\title{Star-Forming Galaxies Significantly Contribute to the Isotropic Gamma-Ray Background}
\author{Tim Linden}
\email{linden.70@osu.edu}
\affiliation{Center for Cosmology and AstroParticle Physics (CCAPP), and \\ Department of Physics, The Ohio State University Columbus, OH, 43210 }

\begin{abstract}
The origin of the isotropic gamma-ray background (IGRB) --- the portion of the extragalactic gamma-ray sky that is not resolvable into individual point sources --- provides a powerful probe into the evolution of the high-energy universe. Star-forming galaxies (SFGs) are among the most likely contributors to the IGRB, though their contribution is difficult to constrain because their flux distribution is dominated by numerous faint sources. We produce a novel joint-likelihood analysis of the $\gamma$-ray emission from 584 SFGs, utilizing advanced statistical techniques to compare the distribution of low-significance excesses against the non-Poissonian $\gamma$-ray background fluctuations. We first examine the theoretically well-motivated relationship between the far-IR and $\gamma$-ray luminosities of SFGs, utilizing a model where the $\gamma$-ray luminosity is given by \mbox{log$_{10}$(L$_\gamma$/(erg s$^{-1}$)) =  $\alpha$~log$_{10}$(L$_{IR}$/(10$^{10}$L$_\odot$)) + $\beta$.} We calculate best-fit parameters \mbox{$\alpha$~=~1.18~$\pm$~0.15}, \mbox{$\beta$~=~38.49~$\pm$~0.24}, with a log-normal dispersion in this relationship given by \mbox{$\sigma$~=~0.39~$\pm$~0.12.} The best-fit values of $\alpha$ and $\beta$ are consistent with previous studies. We find a larger dispersion in the far-IR to $\gamma$-ray correlation than previous studies. This dispersion is significant at the level of 5.7$\sigma$. These results imply that SFGs significantly contribute to the IGRB, producing between 61.0$^{+30.2}_{-18.3}\%$ of the total IGRB intensity above an energy of 1~GeV. Along with recent works, this strongly indicates that multiple source classes provide comparable contributions to the IGRB intensity. We discuss the implication of these results for the interpretation of the IceCube neutrinos.
\end{abstract}

\maketitle

\section{Introduction}
As observed from Earth, the brightest $\gamma$-ray source in the universe is the diffuse emission observed throughout the interstellar medium of the Milky Way. These $\gamma$-rays are produced by cosmic rays interacting with interstellar gas and radiation via hadronic, bremsstrahlung, and inverse-Compton processes. An equivalent process occurs in all galaxies. The intensity of this diffuse $\gamma$-ray emission is thought to be proportional to the galactic supernova rate, as supernovae dominate the cosmic-ray energy budget. There are several complications, most importantly the degree to which any given galaxy is ``calorimetric" to cosmic-rays -- that is, the extent to which the entirety of the cosmic-ray energy budget is converted to radiation (including $\gamma$-rays, radio, or neutrinos) before the cosmic-rays escape. While the majority of galaxies in our universe produce a $\gamma$-ray flux far below the sensitivity of current instruments, a handful of galaxies have been individually detected. These include several of our closest neighbors (LMC, SMC and M31), as well as some of the most intense nearby star-forming galaxies (NGC 1068, NGC 3034 (M 82), NGC 4945, and Arp 220)~\citep{Acero:2015hja,  Peng:2016nsx}. 

The sum of the emission from all galaxies in our universe may produce a significant fraction of the extragalactic $\gamma$-ray flux. It is worth noting that extragalactic $\gamma$-ray emission is divided into two components. The first, known as the ``extragalactic $\gamma$-ray background", includes all $\gamma$-rays produced outside of the Milky Way. The extragalactic $\gamma$-ray background is dominated by emission from $\gamma$-ray blazars~\citep{Ajello:2015mfa, Lisanti:2016jub, Zechlin:2016pme}. The second, which is a subset of the first, includes all extragalactic $\gamma$-rays produced by sources too faint to be individually resolved. Anisotropy constraints imply that blazars can produce no more than $\sim$20\% of this emission, known as the ``isotropic $\gamma$-ray background" (IGRB)~\citep{Ackermann:2012uf, Fornasa:2016ohl}. It is the origin of the latter component that we will study in this paper. The intensity and spectrum of the IGRB has been well-quantified~\citep{2010PhRvL.104j1101A, Ackermann:2014usa}. 

Understanding the IGRB would provide significant insight into the energetics of the high-energy universe~\citep{Fornasa:2015qua, Massaro:2015gla}. In particular, the source that generates the IGRB is likely to produce the high-energy neutrinos observed by IceCube~\citep{Hooper:2016jls}. Possible contributors to the IGRB include blazars~\citep{DiMauro:2014wha}, radio galaxies~\citep{Calore:2013yia, DiMauro:2014wha, Hooper:2016gjy}, star-forming galaxies (SFGs)~\citep{1976MNRAS.175P..23S, Pavlidou:2002va, Thompson:2006qd, Lacki:2012si, 2012ApJ...755..164A, Tamborra:2014xia}, or exotic sources like dark matter annihilation~\citep{Cholis:2013ena, Calore:2013yia, Ackermann:2015tah}

The total emission from cosmic-ray interactions in galaxies is thought to be dominated by those currently undergoing moderate to intense star formation activity. Since star-formation can be traced through far-IR observations (FIR, hereafter considered to be the total 8---1000 $\mu$m emission), models for $\gamma$-ray emission from SFGs have correlated the observed $\gamma$-ray luminosity of each galaxy with the well-measured FIR luminosity. This FIR to $\gamma$-ray relation could then be extrapolated to dim galaxies to determine the total contribution of SFGs to the IGRB. Current models assume a power-law relationship between the $\gamma$-ray and IR luminosities that can be expressed as:

\begin{equation}
\label{eq:correlation}
{\rm log_{10}}\left(\frac{L_{\gamma}}{{\rm erg}~{\rm s}^{-1}}\right) = \alpha~{\rm log_{10}}\left(\frac{L_{IR}}{10^{10} L_\odot}\right) + \beta
\end{equation}
The Fermi-LAT collaboration examined 69 SFGs and obtained best-fit values of $\alpha$=1.17$\pm$0.07 $\beta$~=~39.28$\pm$0.08~\citep{2012ApJ...755..164A} with a source-to-source dispersion of approximately 0.2 dex for $\gamma$-rays in the energy range 100~MeV --- 100~GeV. A recent study of 59 SFGs by~\citep{Rojas-Bravo:2016val} found no new detections (TS=25) compared to those listed in \citep{2012ApJ...755..164A}, and calculated best-fit values $\alpha$=1.12$\pm$0.08, $\beta$~=~-37.9$\pm$0.8\footnote{In \citep{Rojas-Bravo:2016val} the value of $\beta$ was normalized to a FIR luminosity of 1 L$_\odot$, which artificially increases the error on $\beta$, since $\alpha$ and $\beta$ become degenerate.} 

However, these studies investigate only the brightest SFGs, and include only the upper limits from sources that are not detectable at 5$\sigma$ confidence. This significantly degrades the total information available in Fermi-LAT observations. Moreover, these analyses may bias the FIR to $\gamma$-ray correlation in two ways. First, by using only upper limits from non-detected SFGs, these analyses decrease the information extracted \emph{only} from dim SFGs. This potentially makes the correlation systematically biased towards the brightest systems. Second, two of the most important systems for the determination of the FIR to $\gamma$-ray correlation in these analyses are the LMC and SMC. This is due to both their relative proximity (high $\gamma$-ray flux) and low FIR luminosity (which provides them a long lever-arm to constrain the LIR to $\gamma$-ray correlation). However, the close proximity of these systems to the Milky Way is atypical for small galaxies. Additionally, the $\gamma$-ray emission from these systems may not be due to star-formation activity.~\citep{2010A&A...523A..46A, TheFermi-LAT:2015lxa, 2016PhRvD..93f2004C}.

In this study, we analyze the $\gamma$-ray signal coincident with 584 SFGs selected from the IRAS sample~\citep{2003AJ....126.1607S}, and produce a full log-likelihood profile for each SFG as a function of the $\gamma$-ray flux from each sky position. Because many of these SFGs are observed at very low statistical significance, we develop a novel method for estimating the contribution from background fluctuations in the $\gamma$-ray data, creating a joint-likelihood analysis framework that is valid in the presence of large non-Poissonian background fluctuations. Additionally, we allow for system-to-system variation in the correlation between the FIR and $\gamma$-ray luminosities. In this study we neglect the SMC and LMC, but interpret our fit to these systems in the discussion. Utilizing this analysis framework (and marginalizing over all other variables), we find best-fit values of \mbox{$\alpha$~=~1.18~$\pm$~0.15}, \mbox{$\beta$~=~38.49~$\pm$~0.24} and \mbox{$\sigma$~=~0.39~$\pm$~0.12}. We detect dispersion in the FIR to $\gamma$-ray luminosity of SFGs at 5.7$\sigma$. Combining these results with an extrapolation to dim SFGs observed only in the FIR, we find that SFGs contribute a significant fraction of the IGRB, 61.0$^{+30.2}_{-18.3}\%$. This is in slight tension (at the 2$\sigma$ level) with models where SFGs produce the entirety of the IGRB, but also in slight tension (at the 2$\sigma$ level) with analyses showing that radio galaxies dominate the IGRB~\citep{Hooper:2016gjy}. 

\section{Models}
\label{sec:models}

\subsection{Gamma-Ray Data Analysis}
\label{subsec:fermianalysis}
We utilize 584 of the 629 SFGs included in the IRAS sample of bright galaxies~\citep{2003AJ....126.1607S}. We have removed a small number of systems for the following reasons. First, we remove systems observed at a Galactic latitude $|$b$|$~$<$~10$^\circ$, where the bright Galactic diffuse emission produces considerable backgrounds. Second, we have removed the LMC, SMC and M31 from this sample, as they are extended $\gamma$-ray sources that may not be accurately modeled with the point-source analysis technique employed here. Additionally, the FIR luminosities for these systems are significantly smaller than the typical galaxies in our analysis. Detailed $\gamma$-ray analyses of these galaxies exist in the literature~\citep{2010A&A...523L...2A, 2015PhRvD..91j2001B, 2015arXiv150103460P, 2016PhRvD..93f2004C}, and we will compare our results to these systems in the discussion. Third, we remove the system NGC 2146 from our analysis, noting that the system is coincident with a $\gamma$-ray source (3FGL J0707.0+7741). However, in the Fermi 3FGL catalog this source is associated with the nearby BL Lac object NVSS J070651+774137~\citep{Acero:2015hja}, though the association is debated~\citep{Tang:2014dia}. We also produce an analysis which includes the systems M31 and NGC 2146 as $\gamma$-ray bright point sources, finding that they do not significantly affect the results shown here. We note that our IRAS sample does not include Circinus, which has a bright $\gamma$-ray signal that may be produced by either star-formation activity or the Seyfert nucleus of the galaxy~\citep{Hayashida:2013wha}. 

For each of these 584 systems, we calculate the $\gamma$-ray flux using the following technique, which is similar to that employed in~~\citep{Hooper:2016gjy}. Specifically, we employ 84 months of Pass 8 Fermi-LAT data\footnote{MET Range: 239557417-464084557} in the energy range 1~GeV --- 100 GeV using the P8R2\_SOURCE\_V6 event selection criteria. We remove events recorded at a zenith angle exceeding 90$^\circ$, those observed while the instrument is not in survey mode, while the instrumental rocking angle exceeds 52$^\circ$, or while the Fermi-LAT is transiting the South Atlantic Anomaly.

We model the $\gamma$-ray emission in a 14$^\circ\times$14$^\circ$ box surrounding each SFG, binning the dataset into 140$\times$140 angular bins and 10 energy bins logarithmically spaced from 1~GeV to 100~GeV. We first fit the background model (not including the SFG) to the data over the full energy range, allowing the intensity and spectrum of both known $\gamma$-ray point sources and the diffuse $\gamma$-ray emission to float using the default criteria produced using the {\tt make3FGLxml.py} tool. Having fit this background model, we then calculate the improvement in the log-likelihood when a point source is added at the position of the SFG. We initially allow the flux of the SFG to float independently in each energy bin, but then calculate the total improvement in the log-likelihood fit when the SFG is fit to a simple power-law. We test 81 power-law spectra spanning -2.8~$<$~$\gamma$~$<$~-2.0, and for each spectral choice we determine the improvement in the log-likelihood as a function of the $\gamma$-ray flux for each SFG. 

In contrast with previous analyses, we allow the flux of the SFG to be negative, a choice which is non-physical for real sources, but which accounts for the possibility of oversubtraction in the background model. The treatment of negative background fluctuations is critical for this analysis. Because the background model is fit to the data before the SFG is added, the background model is equally likely to undersubtract or oversubtract the $\gamma$-ray data in a given angular bin. SFGs will appear dimmer than expected if they happen to lie coincident with a region of over-subtracted $\gamma$-ray emission. Since we use the Fermi-LAT likelihood algorithm, which utilizes Poisson statistics to compare the number of observed photon counts with the expected number of counts from a given model, mathematical inconsistencies arise in any bin (energy, longitude, latitude) where the negative photon count from SFG model exceeds the positive photon count from the background. We treat this scenario as follows: if the number of observed photons in the bin is 0, we calculate the likelihood by taking the absolute value of the model photon count. This maintains the typical behavior --- the best-fit value for the log-likelihood occurs when the model prediction is also 0, and the log-likelihood fit becomes worse as the model prediction moves away from this value. If the number of observed photons in the bin is non-zero, we add an arbitrarily large error into the log-likelihood calculation for negative model predictions. This also maintains the typical behavior, as a model prediction of exactly 0 photons should also be strongly disfavored if non-zero photons are observed in the bin. We note that this scenario only arises in the highest energy bins, and has no practical implication for our results.

\begin{figure}[tbp]
\centering
\includegraphics[width=.48\textwidth]{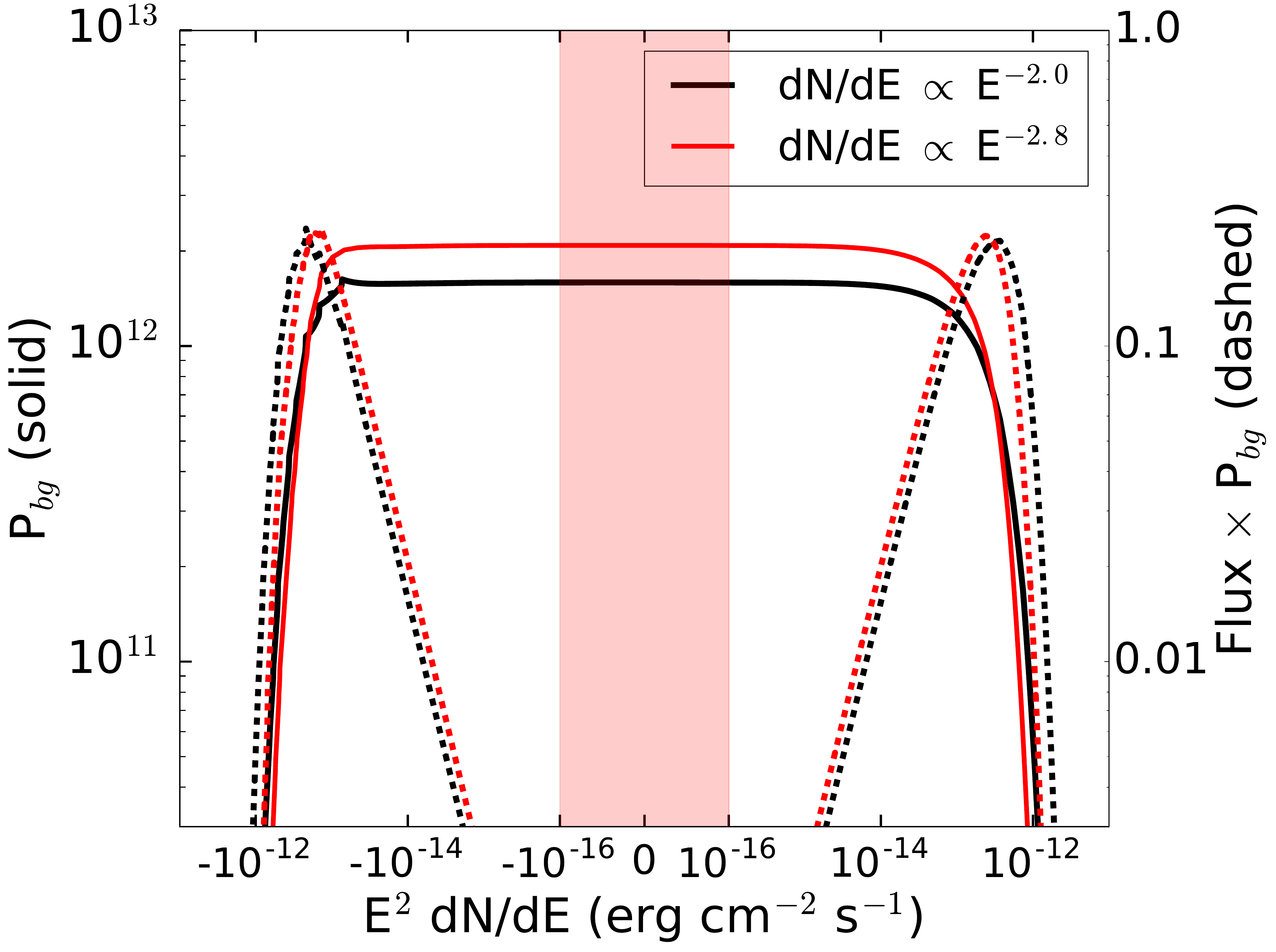}
\caption{The probability of observing a blank sky position with a given $\gamma$-ray flux. The probability distribution is calculated from an ensemble of 1168 background sky positions located at (-$\ell$, $b$) and \mbox{($\ell$, -$b$)} for each SFG in our analysis. The probability distribution (as a function of the $\gamma$-ray flux) for each individual source is taken from the likelihood function obtained by fitting each point source to the Fermi-LAT data. The solid lines show the normalized probability, while the dashed lines show the relative probability of drawing a background fluctuation in a given range $\Delta f$. The red shaded region denotes the portion of the x-axis plotted on a linear scale.}
\label{fig:Pbg}
\end{figure}

\begin{figure}[tbp]
\centering
\includegraphics[width=.48\textwidth]{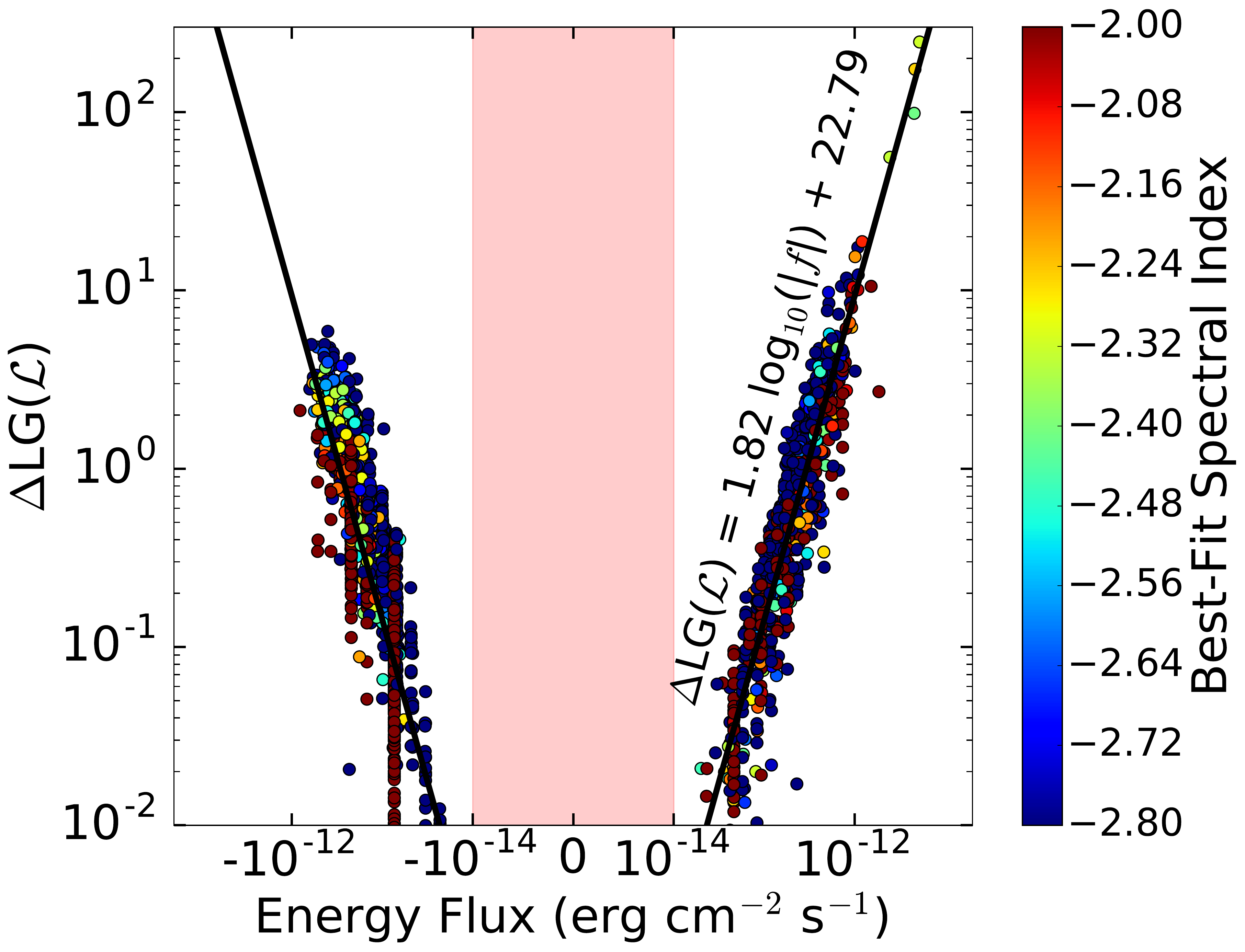}
\caption{The correspondence between the total improvement in the log-likelihood of a point source and the best-fit flux of the source.  The points are taken from all 2336 sky positions (including SFGs, null sky positions, and the positions (-$\ell$, -$b$) used for data verification. We find that the improvement in log-likelihood and the best-fit flux are tightly correlated, as would be expected given that the $\Delta$LG($\mathcal{L})$ is based on the photon count in each energy and angular bin. The best fit relation is given by \mbox{log$_{10}$(LG($\mathcal{L}$)) = 1.818~log$_{10}$($f$) + 22.787}. This highlights the similarity in the approach of using the observed $\gamma$-ray flux distribution at blank sky locations, and the TS distribution from blank sky locations to characterize the background. The red shaded region denotes the portion of the x-axis plotted on a linear scale.} 
\label{fig:LGLflux}
\end{figure}

\subsection{Gamma-Ray to IR Correlation} 

Using the calculated distribution of $\gamma$-ray fluxes, we test the FIR to $\gamma$-ray correlation provided in Equation~\ref{eq:correlation}. We follow the models of~\citep{2012ApJ...755..164A, Hooper:2016gjy} and add an intrinsic scatter in the FIR to $\gamma$-ray correlation, such that the probability of finding a SFG with a given $\gamma$-ray luminosity is given by:

\begin{equation}
\label{eq:luminosity_sigma}
P_{c}\left(L_{\gamma, c}\right) = \frac{1}{2 \pi \sigma^2} {\rm exp} \left( -\frac{{\rm log}(L_{\gamma,c}) - \alpha {\rm log}L_{IR}- \beta}{2 \sigma^2}\right)
\end{equation}

where $L_{\gamma, c}$ is the $\gamma$-ray luminosity of the SFG, and the integral of P$_c$ over all $\gamma$-ray luminosities is normalized to unity. 

\subsection{Background Fluctuations}

As shown in previous studies, the addition of point sources at random positions in the Fermi-LAT sky leads to improvements in the log-likelihood fit that exceed the distribution expected from Poisson statistics~\citep{2014PhRvD..89d2001A, Carlson:2014nra, Hooper:2015ula, 2015PhRvL.115w1301A, Bartels:2015aea, Lee:2015fea, Linden:2015qha}. This significantly affects the calculated flux distribution of low-significance point-sources in Fermi-LAT searches. We must carefully consider the possibility that the flux detected coincident with an SFG results from a fluctuation of the $\gamma$-ray background, rather than from physical emission by the SFG. To account for the effect of background fluctuations in the FIR to $\gamma$-ray correlation, we produce a novel joint-likelihood algorithm that calculates the probability that a fraction (or all) of the $\gamma$-ray flux from any SFG is actually the result of a background fluctuation. The analysis proceeds as follows. 

To determine the distribution of $\gamma$-ray fluxes expected from errors in the background model, we directly evaluate the improvement in the log-likelihood fit for a population of point-sources placed at ``blank" sky locations, where no $\gamma$-ray source is expected. Noting that the brightest diffuse $\gamma$-ray emission is symmetric with respect to the Galactic plane, we choose 1168 blank-sky positions located at the ``mirrored" positions \mbox{(-$\ell$, $b$)} and  \mbox{($\ell$, -$b$)}  for all SFGs at locations \mbox{($\ell$, $b$)} in our analysis. This ensures that our blank sky positions have a similar global spatial distribution as the real SFGs. We also analyze the $\gamma$-ray emission from the blank sky positions located at \mbox{(-$\ell$, -$b$)}, but will reserve these final positions to produce tests of our analysis method shown in the Appendices.

We note that Equation~\ref{eq:luminosity_sigma} can be rewritten as a correlation between the FIR and  $\gamma$-ray fluxes. Specifically, we take:

\begin{equation}
\label{eq:lum_to_flux}
\phi_c  = \frac{1}{4 \pi d_{SFG}^2} L_{\gamma, c}
\end{equation}

for each SFG. Running our analysis on the 1168 blank sky positions, we calculate the fit to the $\gamma$-ray data as a function of the $\gamma$-ray flux, producing a likelihood function for each SFG denoted as LG($\mathcal{L}$($\phi_{bg}$)). We set LG($\mathcal{L}$($\phi_{bg}$))~=~0 at the best-fit value of the $\gamma$-ray flux. The probability of observing a $\gamma$-ray flux $f$ at a given sky position can be computed by taking the exponential of the log-likelihood function. This is similar (in spirit) to producing a histogram of background counts to determine the distribution of noise in an observation --- except the error bars on the $\gamma$-ray flux are large, so each point in the blank sky analysis contributes continuously to many bins in the histogram.  Examining the combined population of blank sky positions, the probability that a given $\gamma$-ray flux will be observed from background fluctuations in the Fermi-LAT data is given by:

\begin{equation}
\label{eq:p_bg}
P_{bg}(\phi_{bg}) = \frac{1}{N}\sum_{i} A_i ~{\rm exp}\left(-{\rm LG}(\mathcal{L}(\phi_{bg}))\right)
\end{equation}

where $\phi_{bg}$ is the background $\gamma$-ray flux, $\Delta$LG($\mathcal{L}$($\phi_{bg}$)) is the likelihood of the fit to a given blank-sky position $i$ at a flux value of $\phi_{bg}$. The sum is taken over all N=1168 blank sky positions in our analysis, and A$_i$ is a set of normalization constants set such that the integral of ${\rm exp}\left(-{\rm LG}(\mathcal{L}(\phi_{bg})\right)$ over all $\phi_{bg}$ is unity for each blank sky position.

In Figure~\ref{fig:Pbg} we show the resulting function P$_{bg}$ calculated through our analysis of 1168 blank sky positions using the same analysis framework described in Section~\ref{subsec:fermianalysis} to determine the $\gamma$-ray flux from physical SFGs. We note that the value of P$_{bg}$ is calculated over many orders of magnitude in the observed flux, while the probability of observing a $\gamma$-ray flux in a specified range is given by P$_{bg}$($f$)$\Delta f$. Thus, we also show the value of P$_{bg}$ re-weighted by the flux to illustrate the relative probability of drawing a given $\gamma$-ray flux from a blank sky position.

Using P$_{bg}$, we can calculate the probability that a point-source with a given $\gamma$-ray flux is consistent with fluctuations due to background mismodeling. A physical SFG with a best-fit $\gamma$-ray flux near the maximum value of P$_{bg}$ would have a high probability of being explained as a background fluctuation, while an SFG with a $\gamma$-ray flux near the tails of the P$_{bg}$ distribution would have a small chance of being explained by background mismodeling. 

It is possible that a background fluctuation can explain some, but not all, of the $\gamma$-ray flux from a true point source. That is, if a point source has a likelihood function given by LG($\mathcal{L}$($\phi$)), the probability that the source is consistent with a given $\gamma$-ray flux is given by ${\rm exp}\left(-{\rm LG}(\mathcal{L}(\phi)\right)$. The probability that this source can be explained by a background fluctuation can be calculated by multiplying the probabilities of each distribution, that is: ${\rm exp}\left(-{\rm LG}(\mathcal{L}(\phi)\right)$ $\times$ P$_{bg}(\phi)$. This value may be maximized at a flux $\phi$ that is not equal to the best-fit flux for the point source. Any remaining flux may be explained as an unusually bad error in the fitting of the $\gamma$-ray data, or may be explained by physical emission stemming from a true point source at the given sky location. Formally, we can combine Equations~\ref{eq:luminosity_sigma}~through~\ref{eq:p_bg} to calculate the probability that the $\gamma$-ray flux from SFGs can be explained via the FIR to $\gamma$-ray correlation as:

\begin{multline}
\label{eq:theequation}
P(\alpha, \beta, \sigma) = \prod_j \int_{-\infty}^\infty \int_0^\infty {\rm exp}(-{\rm LG}(\mathcal{L}(\phi_c + \phi_{bg}))) \times \\  \times P_{bg}(\phi_{\gamma,bg})P_{c}\left(\phi_c, \alpha, \beta, \sigma \right) \; {\rm d\phi_c}\, {\rm d \phi_{bg}}
\end{multline}

This formalism determines the probability that an SFG with a given likelihood function is explained by some additive combination of a background fluctuation and a true source flux predicted by the FIR to $\gamma$-ray relation. The term ${\rm exp}(-{\rm LG}(\mathcal{L}(\phi_c + \phi_{bg}))$ is independently maximized at the best-fit flux of the SFG under consideration, while the P$_{bg}$ and P$_c$ terms provide the probability that the source fluxes $\phi_c$ and $\phi_{bg}$ are consistent with the expectations from the background model and the FIR to $\gamma$-ray correlation, respectively. Note that $\phi_{bg}$ can be negative, as it includes terms due to the mismodeling of the $\gamma$-ray background. On the other hand, $\phi_{c}$ cannot be negative, as it represents the true $\gamma$-ray flux from the SFG under investigation. The total probability of the fit to the FIR to $\gamma$-ray correlation is the product of the probability over all 584 SFGs. We maximize P($\alpha$, $\beta$, $\sigma$) to determine the best-fit FIR to $\gamma$-ray correlation for the SFG population.

We note in Figure~\ref{fig:Pbg} that the distribution of P$_{bg}$ varies based on the calculated spectrum of the SFG under investigation. In this analysis, we calculate 81 functions P$_{bg, i}$ spanning the spectral indices -2.8~$<$~$\gamma$~$<$-2.0. For each SFG, we calculate the best-fit spectral index from a scan of the LG($\mathcal{L}$) parameter space (ignoring the possibility of background fluctuations), and then utilize the likelihood function and the function P$_{bg, i}$ that correspond to the best-fit SFG spectrum. We find that this choice, compared to the choice of a single spectral index for our entire analysis, has a negligible effect on our results.

\subsection{Contribution to the IGRB}

\begin{figure*}[tbp]
\centering
\includegraphics[width=.49\textwidth]{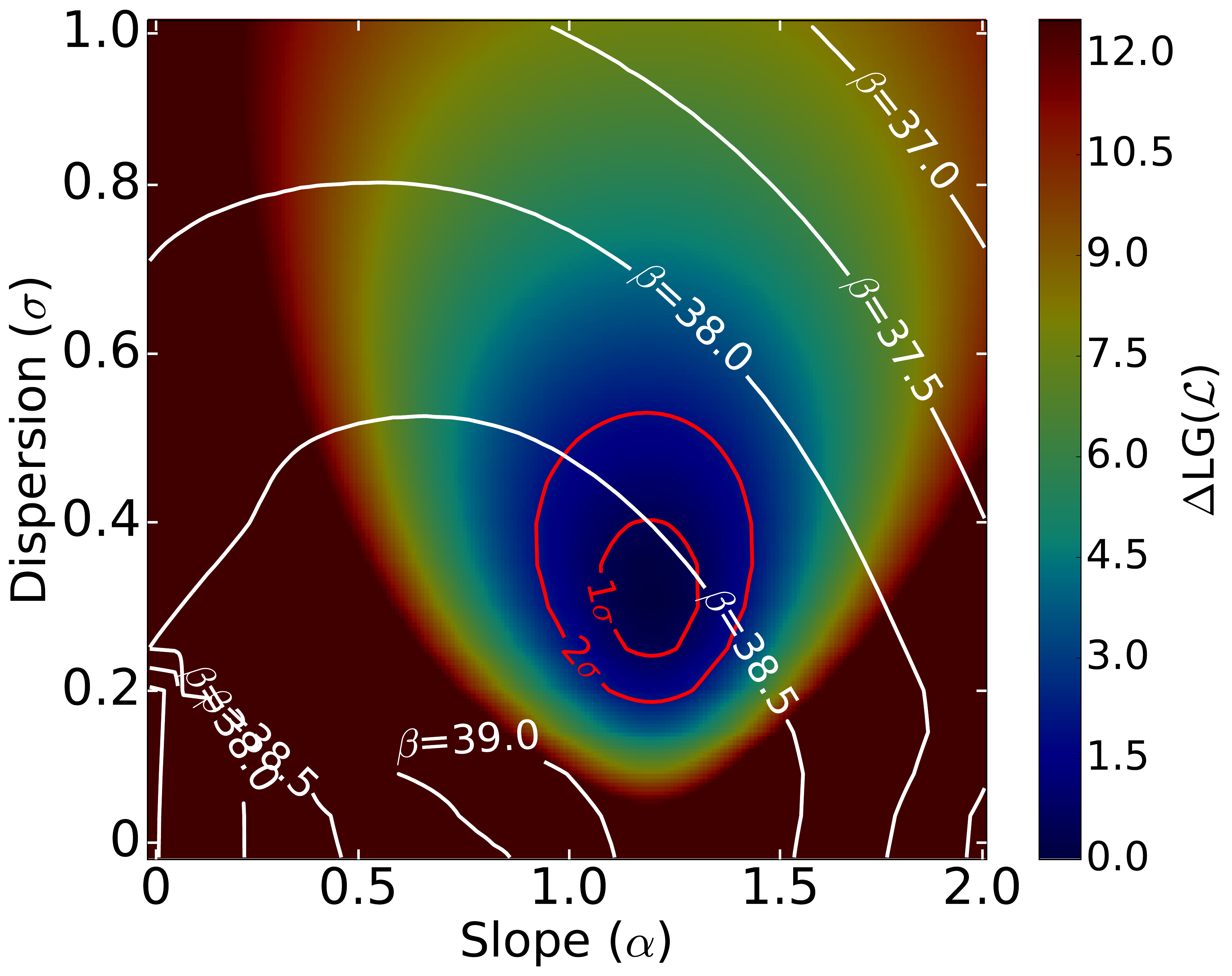}
\includegraphics[width=.49\textwidth]{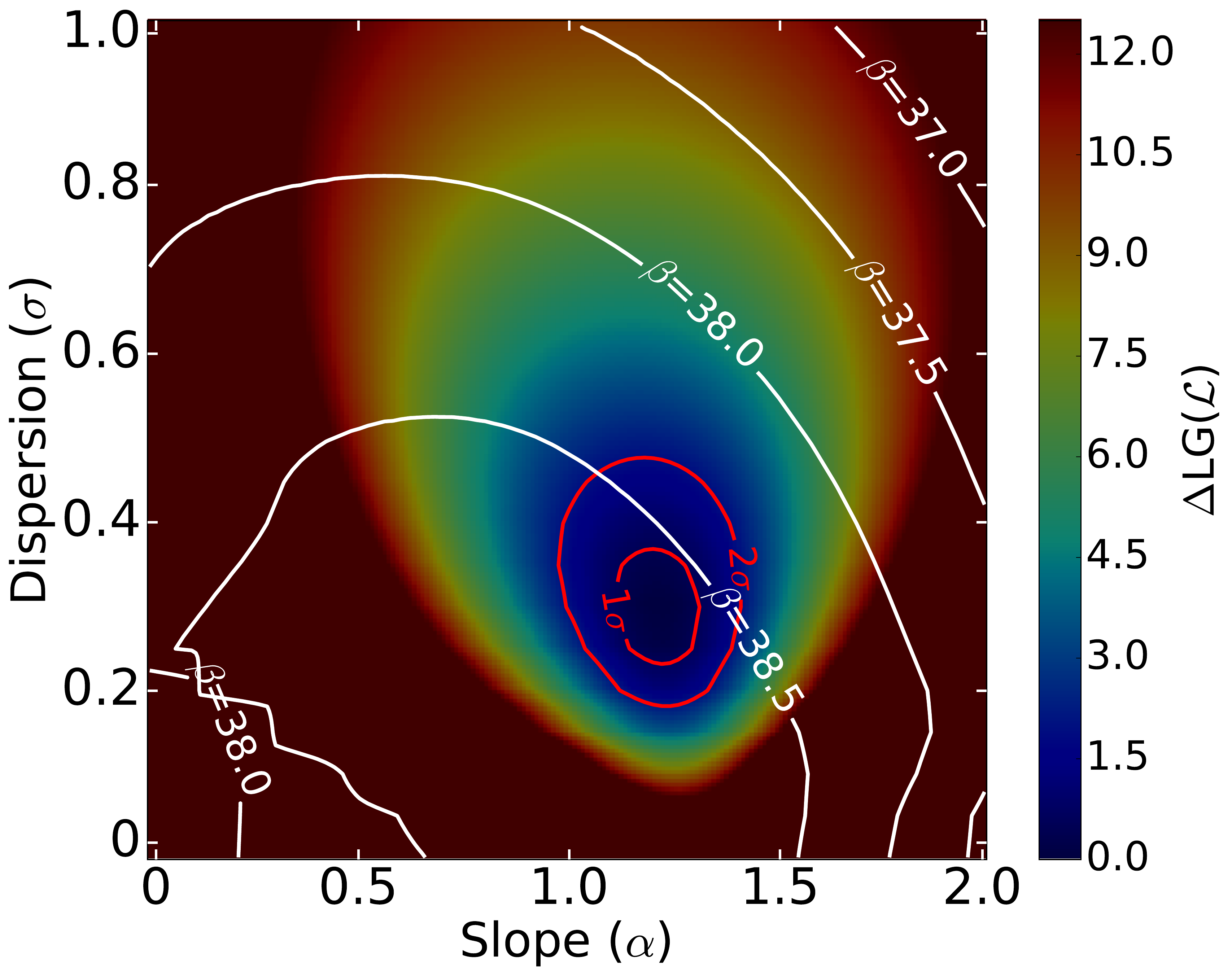}

\caption{The best fit correlation between the FIR and $\gamma$-ray luminosities (1---100~GeV) for the population of 584 SFGs in our sample in a model that excludes (left, default) or includes (right) the extended $\gamma$-ray sources NGC 224 and NGC 2146. The 1$\sigma$ and 2$\sigma$ uncertainties in the correlation are given by the red error ellipses, while the white contours provide the best fit value of $\beta$ for each point $\alpha$, $\sigma$. We find statistically significant evidence for a dispersion in the FIR to $\gamma$-ray correlation. The value $\sigma$~=~0.0 is rejected at 5.7$\sigma$ significance in our default model. The best fit parameters for the correlation are $\alpha$~=~1.18~$\pm$~0.15, $\beta$~=~38.49~$\pm$~0.24 and $\sigma$~=~0.39~$\pm$~0.12 (each error bar is calculated by marginalizing over the other parameters). We note that these uncertainties, in particular those of $\beta$ and $\sigma$, are partially degenerate. } 
\label{fig:sfgplot}
\end{figure*}

Using the calculated FIR to $\gamma$-ray relation, we can calculate the total $\gamma$-ray contribution to the IGRB by extrapolating to low FIR luminosities. For this, we closely follow the calculation of~\citep{Tamborra:2014xia}, using the FIR flux determination normalized to~\citep{Gruppioni:2013jna}. We repeat a few details of the calculation below for clarity, but refer the reader to these works for a detailed explanation of the technique.

We adopt a model for the FIR luminosity function given by~\citep{Gruppioni:2013jna} as:

\begin{multline}
\Phi_{IR, X}(L_{IR}, z){\rm dlog} L_{IR} = \Phi^*_{IR, X}(z) \left(\frac{L_{IR}}{L^*_{IR, X}(z)}\right)^{(1-\alpha_{IR, X})}\\
\times {\rm exp}\left[-\frac{1}{2\sigma^2_{IR,X}}log^2\left(1 + \frac{L_{IR}}{L^*_{IR,X}(z)}\right)\right] {\rm dlog} L_{IR}
\end{multline}

where the subscript X distinguishes the redshift evolution of normal galaxies, starburst galaxies, and starburst galaxies with possible active galactic nuclei. In the observational work of~\citep{Gruppioni:2013jna}, the values of $\Phi^*_{IR, X}$ and L$^*_{IR, X}$ are fit to the redshift bin z = 0.0 --- 0.3. In \citep{Tamborra:2014xia}, these values are recalculated to provide best fit values at redshift z=0 to simplify the calculation. However, the reported values in \citep{Tamborra:2014xia} appear to overestimate the data recorded by \citep{Gruppioni:2013jna} in the redshift range 0.0---0.3. In this work, we calculate best-fit values of $\Phi^*$ and L$^*$ such that the total emission from the redshift range z=0.0 --- 0.3 is equivalent to the data recorded by \citep{Gruppioni:2013jna}. Specifically we calculate best-fit values of log$_{10}$(L$^*$)~=~\{9.46, 11.02, 10.57\}~L$_\odot$  and log$_{10}$($\Phi^*$)~=~\{-2.08, -4.74, -3.25\} Mpc$^{-3}$ for the ``Normal Galaxy", ``Starburst Galaxy" and ``Star-Forming AGN"  components respectively. We find that these choices decrease the total $\gamma$-ray contribution to the IGRB by approximately 20\%, compared to using values from \citep{Tamborra:2014xia}. Utilizing this distribution of FIR luminosities, we calculate the total $\gamma$-ray intensity as:

\begin{multline}
\label{eq:integral}
I(E_\gamma) = \int_0^{z_{max}}dz \int_{L_{\gamma, min}}^{L_{\gamma, max}} \frac{dL_\gamma}{{\rm LG}(10)L_\gamma} \frac{d^2V}{d\Omega dz} \times \\
\times \sum_X \Phi_{\gamma, X}(L_\gamma, z) \frac{dF_{\gamma, X}(L_\gamma, (1+z)E_\gamma, z)}{dE_\gamma}e^{-\tau(E_\gamma, z)}
\end{multline}

where d$^2$V/d$\Omega$dz is the comoving volume, $\tau$ is the optical depth of $\gamma$-rays at a given energy and redshift \citep{Gilmore:2011ks}, dF$_{\gamma, X}$/dE$_\gamma$ is the $\gamma$-ray spectrum, for which we use a constant power-law $\alpha$=-2.32 (based on our analysis of the brightest SFGs described later). Finally, we note that Equation~\ref{eq:correlation} is used to replace the unknown $\gamma$-ray flux distribution with the known FIR flux distribution, utilizing the fact that:

\begin{equation}
\Phi_{\gamma, X}(L_\gamma, z) {\rm dlog}L_\gamma = \Phi_{IR, X}(L_{IR}, z)  {\rm dlog}L_{IR}
\end{equation}

We note that the above model does not include dispersion in the FIR to $\gamma$-ray correlation. To determine the contribution of SFGs to the IGRB for a given value of the dispersion in the FIR to $\gamma$-ray relation, we first calculate the mean $\gamma$-ray luminosity of SFGs in this model as a function of the FIR luminosity. We then calculate the total SFG contribution to the IGRB utilizing this new mean value of $\beta$, assuming no additional dispersion. We note that this provides mathematically identical results in the limit that a large number of SFGs contribute to the IGRB --- that is, so long as the log-normal distribution of $\gamma$-ray fluxes is fully explored by the SFG population. Otherwise, the mean contribution of SFGs to the IGRB will stay the same, but the error bars will be underestimated. Later in the paper we justify this assumption, finding that the majority of the IGRB intensity is produced by millions of systems with similar $\gamma$-ray fluxes. 

We make two theoretically motivated cuts on the $\gamma$-ray flux in our models. First, we ignore the $\gamma$-ray flux produced by any system that would be detectable as a $\gamma$-ray point source in our analysis. This is necessary to make a proper comparison with the IGRB, which includes only the emission that remains after all detectable point-sources are removed. Removing these sources is mathematically equivalent to setting a minimum bound on the redshift integral that is dependent on the $\gamma$-ray luminosity of the SFG. To determine which point sources would be detectable at the TS=25 level (used in the 3FGL catalog), we calculate the correspondence between the maximum improvement in the log-likelihood for a point source and the best-fit flux of that point source. We show our results in Figure~\ref{fig:LGLflux}, and find that these parameters are tightly correlated. Using this relationship, we do not include a contribution from any system with a flux exceeding 1.17~$\times$10$^{-12}$~erg~cm$^{-2}$~s$^{-1}$ over 1~GeV, which would correspond to TS~=~25  in our analysis. We find that this cut decreases the SFG contribution to the IGRB by less than 1\%.

\begin{figure}[tbp]
\centering
\includegraphics[width=.48\textwidth]{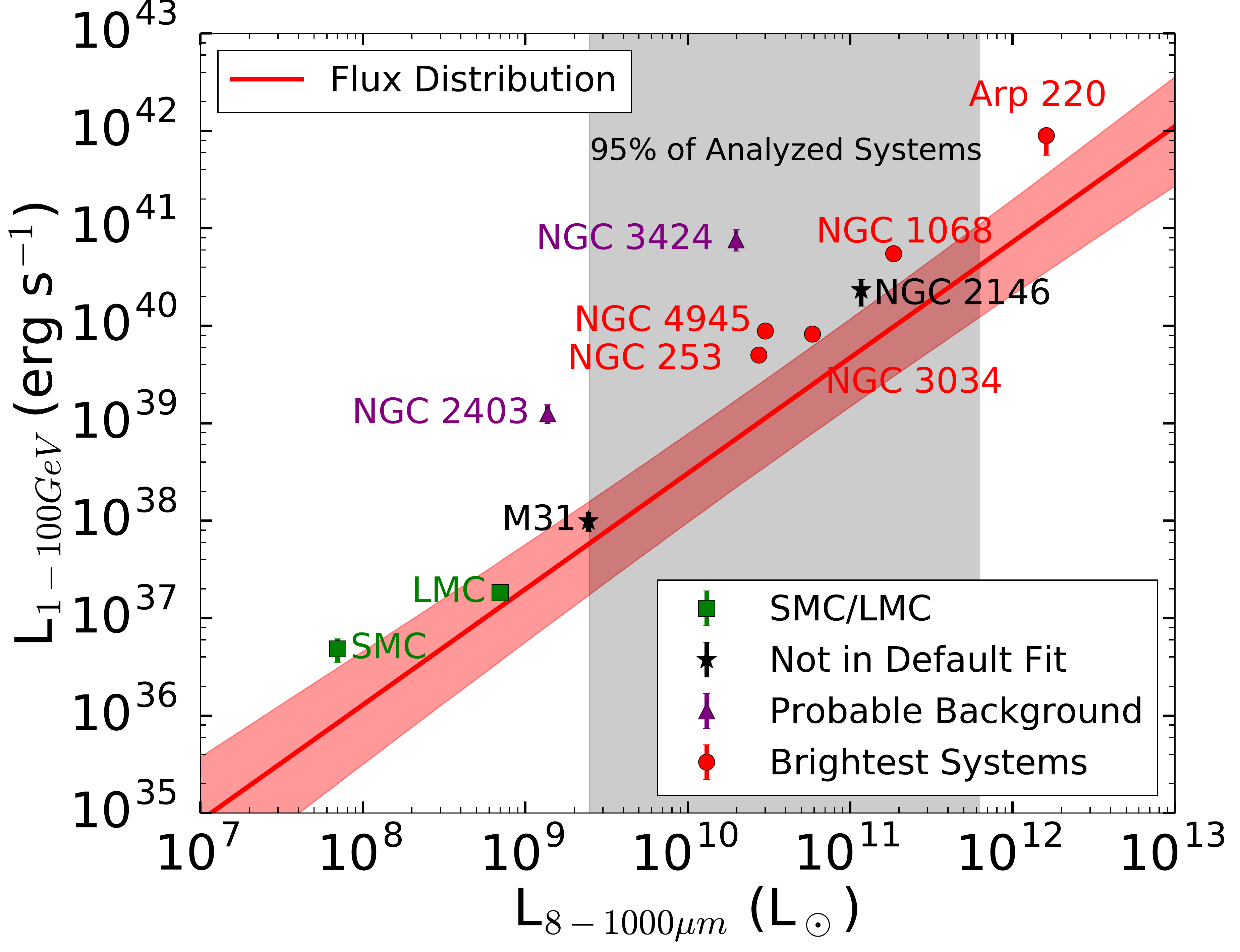}
\caption{The $\gamma$-ray luminosity distribution of SFGs in our sample (red with 1$\sigma$ error bars) as a function of the FIR luminosity of each SFG. Data is shown for the brightest $\gamma$-ray SFGs in our model (as well as for M31 and NGC 2146 which are excluded from our default analysis). The SMC and LMC are also shown, using the best-fit flux calculated by \citep{2010A&A...523A..46A, 2010A&A...512A...7A} and translated to a flux in the 1---100~GeV range. It is clear that the SFGs that are detected in $\gamma$-ray data are systematically bright compared to the average SFG population. The gray shaded region encompasses 95\% of the SFGs in our sample.} 
\label{fig:lir_lgamma}
\end{figure}

\begin{figure}[tbp]
\centering
\includegraphics[width=.48\textwidth]{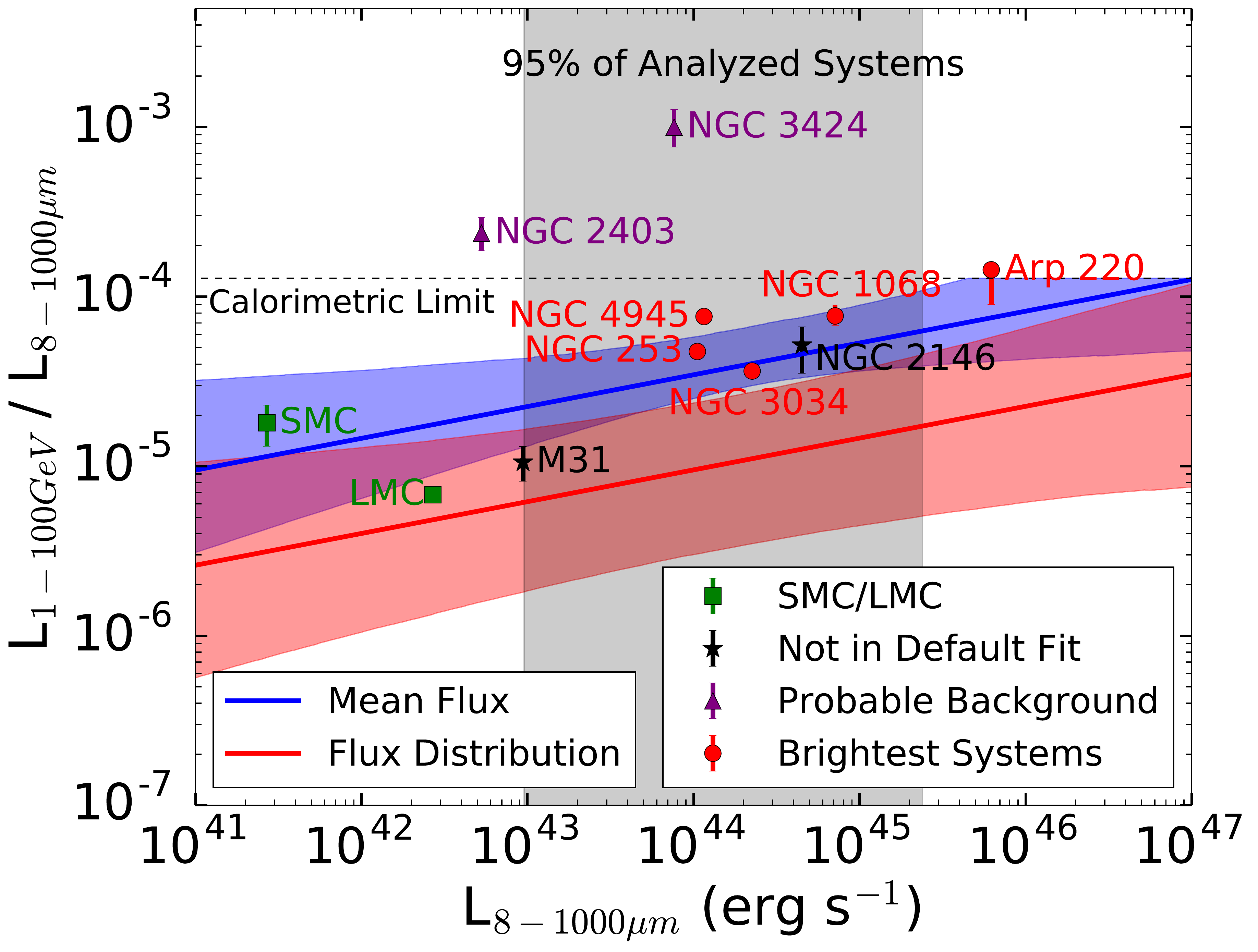}
\caption{The ratio of the best-fit $\gamma$-ray luminosity compared to the FIR luminosity of SFGs in our model, as a function of the FIR luminosity. This is equivalent to Figure~\ref{fig:lir_lgamma} with the y-axis divided by the x-axis. The flux distribution (with 1$\sigma$ error bars) of SFGs is again plotted in red. The mean flux of SFGs in our model (with 1$\sigma$ error bars) is plotted in blue. The mean flux exceeds the median flux of SFGs because the SFG luminosity is allowed to vary via a log-normal distribution -- which causes the total flux to be dominated by the brightest systems. The calorimetric limit is set at 1.28$\times$10$^{-4}$, as described in the text. We do not allow the average $\gamma$-ray flux in any model to exceed the calorimetric limit for any value of L$_{IR}$. This explains the cutoff in the upper-limit of the mean flux at high IR luminosities. } 
\label{fig:lir_lgamma_frac}
\end{figure}

Second, we do not allow the mean $\gamma$-ray luminosity from SFGs in our model to be super-calorimetric for any value of the FIR luminosity. This scenario can occur during the extrapolation of the FIR to $\gamma$-ray correlation up to high FIR luminosities in cases where $\alpha$ is large. While a given SFG may be super-calorimetric due to time variability in the FIR to $\gamma$-ray correlation, the average flux from such systems should not be above the calorimetric limit, due to energy conservation. We reduce the $\gamma$-ray emission from these systems to the calorimetric limit, utilizing the relationship calculated in ~\citep{Lacki:2010vs}, but setting the maximum $\gamma$-ray luminosity to be 1.28~$\times$10$^{-4}$ of the FIR luminosity in the energy range 1---100~GeV. This value is slightly lower than calculated in ~\citep{Lacki:2010vs}, which considers the total $\gamma$-ray emission in the range 0.1---1000~GeV. We convert between these models assuming a best-fit spectral index for SFG, which we derive later. We find that this cut decreases the total SFG contribution to the IGRB by approximately 10\%.

\section{Results}

\subsection{Analysis of Individual SFG Properties}
Our analysis confirms previous results indicating that the systems NGC 253, NGC 1068, NGC 3034 and NGC 4945 are $\gamma$-ray bright SFGs. Utilizing the distribution of P$_{bg}$ to translate the likelihood function of each point source the statistical significance of the source (taking into account the possibility of background mismodeling), we can determine that the $\gamma$-ray emission observed from each SFG is statistically significant at confidence levels of: 7.00$\sigma$ (NGC 253), 4.42$\sigma$ (NGC 1068), 8.06$\sigma$ (NGC 3034), and 7.10$\sigma$ (NGC 4945). Because the distribution P$_{bg}$ takes into account the possibility of background fluctuations, we note that all of these sources are highly significant. We discuss the utilization of P$_{bg}$ to determine the statistical significance of sources in Appendix~\ref{sec:statsignificance}.

In addition to the SFGs identified in the 3FGL, our analysis identifies the systems NGC 2403 and NGC 3424 to have $\gamma$-ray emission with a statistical significance that exceeds the TS~=~25 barrier typically used for source identification, with TS values of 37.5 and  30.9 respectively. However, by using P$_{bg}$ to determine the possibility that these systems are the result of background fluctuations, we find that the statistical significance of these systems is only 2.55$\sigma$ and 2.36$\sigma$, respectively.  Given that we analyze 584 SFGs, it is reasonable that two false detections would be found at these significances. We calculate the TS for Arp 220 to be 23.4, which just barely falls short of the TS~=~25 threshold. Compared to the analysis of \citep{Peng:2016nsx}, we note that our higher threshold energy of 1~GeV (important for the systematic study that follows) is likely to make us less sensitive to emission from Arp 220. We thus consider these observations to be compatible.

Determining the spectra of SFGs is harder, as systems observed at low-statistical significance have very poorly defined $\gamma$-ray spectra. Using the four brightest systems in our analysis (NGC 253, NGC 1068, NGC 3034 and NGC 4945), we calculate the average $\gamma$-ray spectrum above 1~GeV to follow a power-law with a spectral index of $\gamma$~=~-2.32$\pm$0.06. These four systems are all individually best-fit with power-law spectra between -2.2 and -2.4.  There are no statistically significant signs of dispersion in the average $\gamma$-ray spectrum. In what follows we will take $\gamma$~=~-2.32 to be the nominal spectrum for the entire SFG population, but we note that there are large systematic uncertainties in this relationship. Since the calculated $\gamma$-ray spectrum is critical to determine the total $\gamma$-ray intensity at high-energy, there is significant uncertainty in the contribution to the IGRB above several tens of GeV. 

\begin{figure}[tbp]
\centering
\includegraphics[width=.49\textwidth]{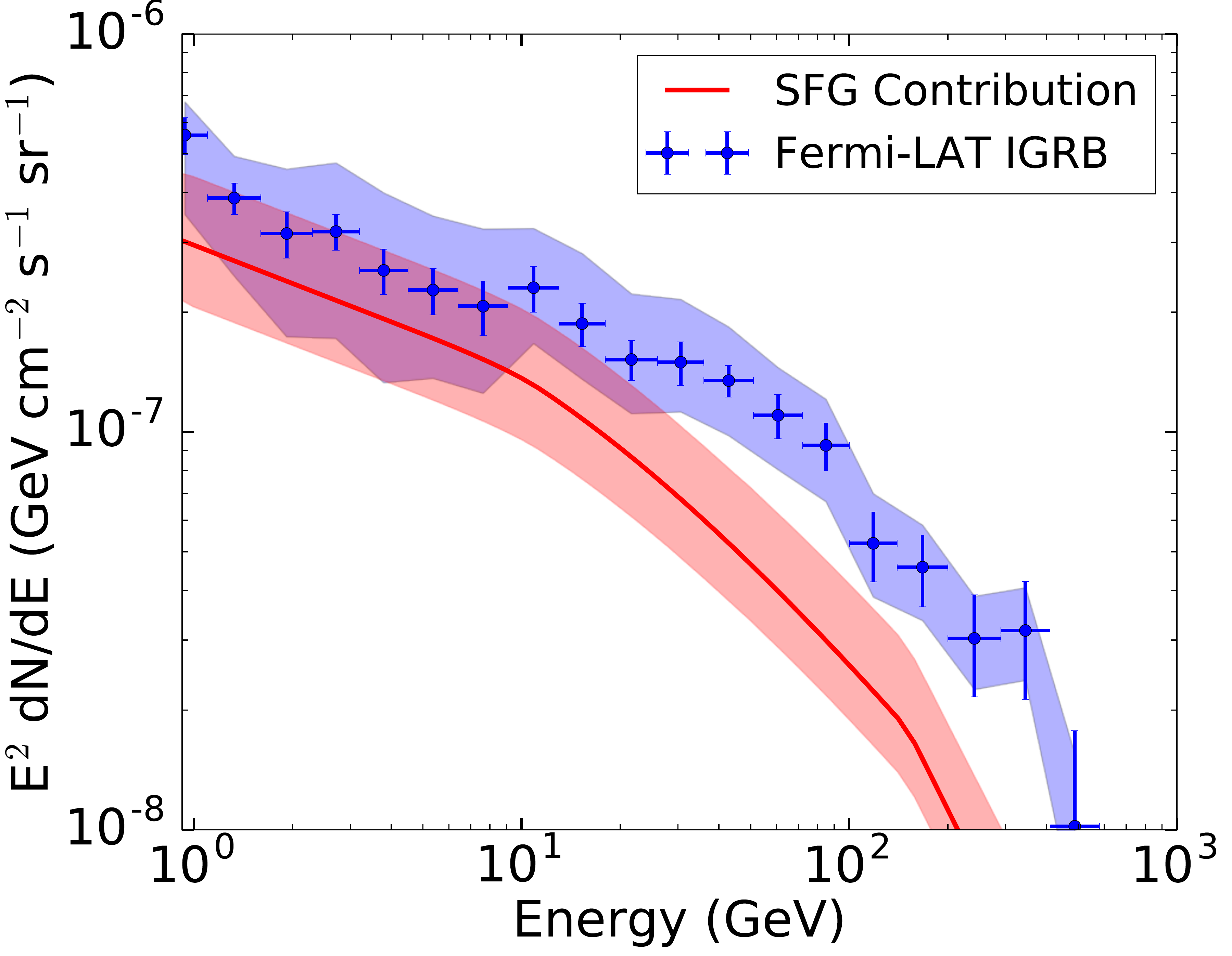}
\caption{The contribution of SFGs to the IGRB, as calculated in~\citep{Ackermann:2014usa}. The blue-error bars refer to the statistical errors in the determination of the IGRB, while the blue error bands relate to systematic uncertainties in the subtraction of Galactic foreground emission. The solid red line denotes the mean contribution of SFGs calculated as the average of 10,000 points in the ($\alpha$, $\beta$, $\sigma$) parameter space, weighted by their fit to the $\gamma$-ray to IR correlation produced in our model. The shaded red band relates to the 1$\sigma$ uncertainties in the SFG contribution to the IGRB. Note that the intensity of the IGRB is not taken into account in models of the SFG, and there is no correlation which prohibits the SFG from overproducing the IGRB in these models.} 
\label{fig:igrbplot}
\end{figure}

\subsection{FIR to $\gamma$-ray Correlation}
In Figure~\ref{fig:sfgplot} we show the best fit values of $\alpha$, $\beta$, and $\sigma$ in the FIR to $\gamma$-ray correlation for SFGs. We show the best-fit parameters for our default scenario where the systems NGC 224 and NGC 2146 are ignored (left), and in an alternative scenario where NGC 224 is modeled as a point-source and NGC 2146 is modeled under the assumption that the nearby blazar produces no $\gamma$-ray emission (right). We find the results in each case to be similar and note three key results. First, we find strong evidence for a FIR to $\gamma$-ray correlation, statistically significant at the level of 12.6$\sigma$. Second, we find statistically significant evidence for dispersion in this correlation. Models with no dispersion ($\sigma$~=~0.0), are excluded at 5.7$\sigma$ confidence. In Appendix~\ref{sec:injectedsignal} we provide strong evidence that this dispersion is intrinsic to the SFG population, and is not a result of mismodeling  the $\gamma$-ray background. While the best-fit dispersion is $\sigma$~$\approx$~0.35, significantly larger dispersions are not strongly excluded by the data. We identify a significantly larger dispersion compared to the analysis of \citep{2012ApJ...755..164A}, which finds a dispersion of only $\sigma$~=~0.2, and notes that this is an upper-limit, as the dispersion could be reduced if errors in the $\gamma$-ray flux of each source were taken into account. In this analysis, we take those errors into account, and find that the inclusion of full likelihood profiles for the non-detected systems significantly increases the best-fit value of $\sigma$, as discussed in Appendix~\ref{sec:appendixsubthreshold}.

Third, we find a best fit value of $\alpha$~$\approx$~1.2, compatible with previous studies. While our best-fit value of $\beta$~$\approx$~38.5 is smaller than in previous works, we note that this parameter is degenerate with the degree of dispersion in our model. In particular, since the source-to-source variation is described with a log-normal distribution, the mean SFG luminosity is significantly higher than the median SFG luminosity. The mean SFG luminosity for $\beta$~=~38.5, $\sigma$~=~0.3 would be equivalent to a model with $\beta$~=~38.81 and no dispersion. This result is similar to previous works, after additionally accounting for the fact that the $\gamma$-ray luminosity in this study is calculated in the energy range 1---100~GeV. After marginalizing over each of the other two variables, we find best fit values of $\alpha$~=~1.18~$\pm$~0.15, $\beta$~=~38.49~$\pm$~0.24, and $\sigma$~=~0.39~$\pm$~0.12. We note that these errors would be somewhat smaller if there were no degeneracies between the model parameters. We will discuss the theoretical consequences of dispersion in the FIR to $\gamma$-ray correlation in an forthcoming publication.

In Figures~\ref{fig:lir_lgamma} and \ref{fig:lir_lgamma_frac} we show the corresponding relationship between the FIR luminosity of SFGs in our sample and the calculated $\gamma$-ray luminosity of these systems. To determine the range of $\gamma$-ray luminosities that are compatible with the FIR to $\gamma$-ray correlation, we calculate the $\gamma$-ray emission from 10,000 random points in the ($\alpha$, $\beta$, $\sigma$) parameter space that are within 3$\sigma$ of the best-fit values of each parameter. Four results become immediately apparent: (1) the observed SFGs are systematically bright compared to the median SFG in our model. Detected SFGs are typically 1-2$\sigma$ outliers compared to the median SFG flux, (2) The SMC and LMC lie near the best-fit FIR to $\gamma$-ray correlation, despite not being fit in our models. This lends credence to the extrapolation of the FIR to $\gamma$-ray correlation to systems with low star-formation rates. (3) In Figure~\ref{fig:lir_lgamma_frac} we additionally plot the mean $\gamma$-ray flux from SFGs (with 1$\sigma$ error bars corresponding to the variance based on the 10,000 weighted model fits to the FIR to $\gamma$-ray correlation). We note that the mean flux is systematically higher than the flux of the median SFG. This is due to the log-normal dispersion in the L$_{IR}$ to L$_\gamma$ relationship utilized in our model. (4) The systems NGC 3424 and NGC 2403 lie far off of the FIR to $\gamma$-ray correlation, providing some evidence that these systems either stem from background fluctuations, or are $\gamma$-ray bright for reasons unrelated to their star-formation rate. We note that Equation~\ref{eq:theequation} is resilient to the significant mismodeling of these systems. Since they lie far off of the FIR to $\gamma$-ray correlation, no values of $\alpha$, $\beta$, and $\sigma$ are likely to provide a good fit. Thus, they have only a marginal influence on our best-fit parameters.

\subsection{SFG Contribution to the IGRB}

\begin{figure}[tbp]
\centering
\includegraphics[width=.47\textwidth]{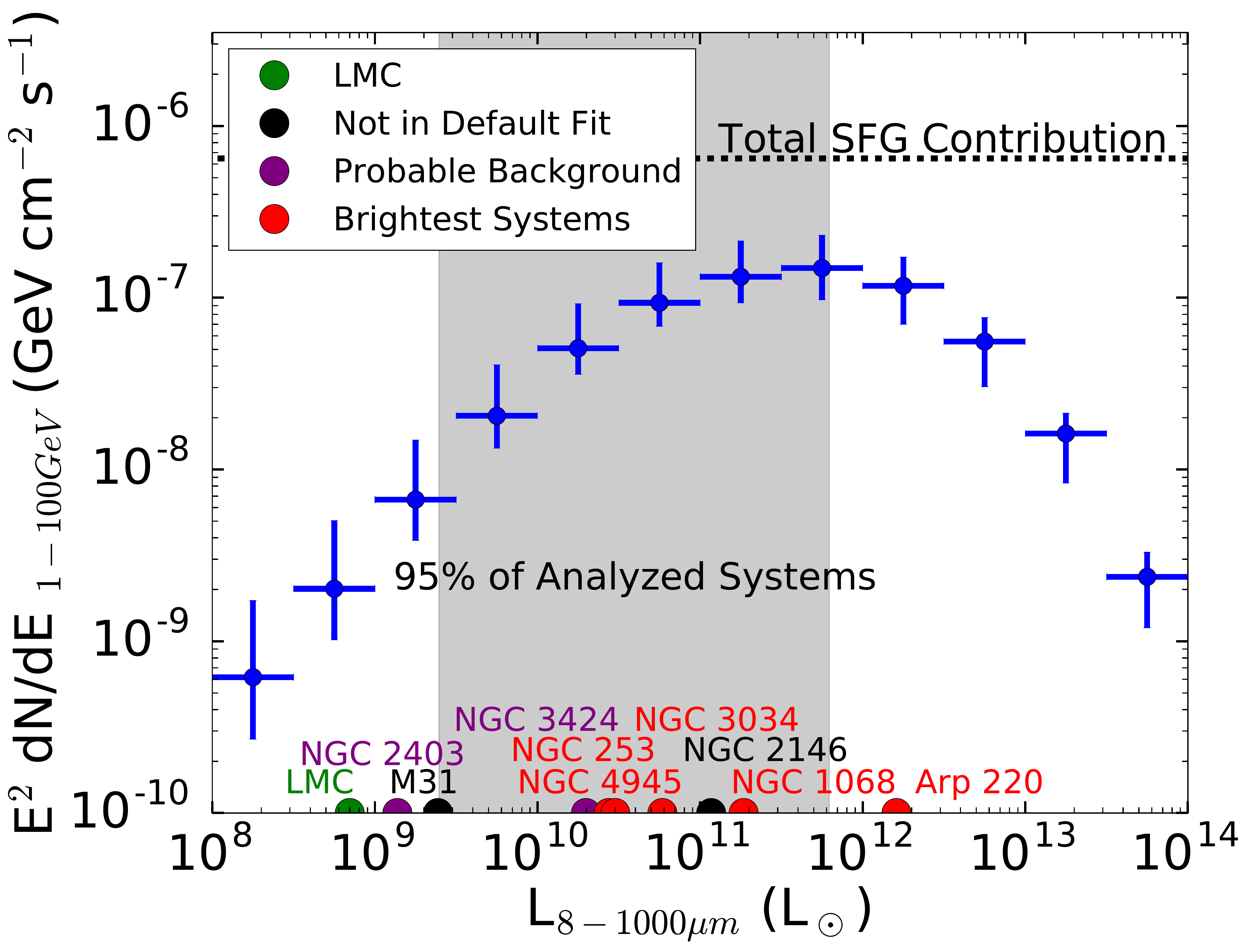}
\caption{The relative contribution of SFGs to the IGRB as a function of their FIR luminosity. We see that the majority of the SFG contribution to the IGRB stems from systems with FIR luminosities between 10$^{11}$ --- 10$^{12}$~L$_\odot$, on par with the brightest SFGs that we observe in the local universe. Systems with low star-formation rates contribute negligibly to the total emission. The FIR luminosity of the systems most important in our analysis are shown at the bottom of the plot (their y-coordinates are arbitrary). We see that the average IGRB contribution stems from systems with FIR luminosities slightly higher than the average system in our sample.} 
\label{fig:lgamma_binned_contribution}
\end{figure}

\begin{figure}[tbp]
\centering
\includegraphics[width=.47\textwidth]{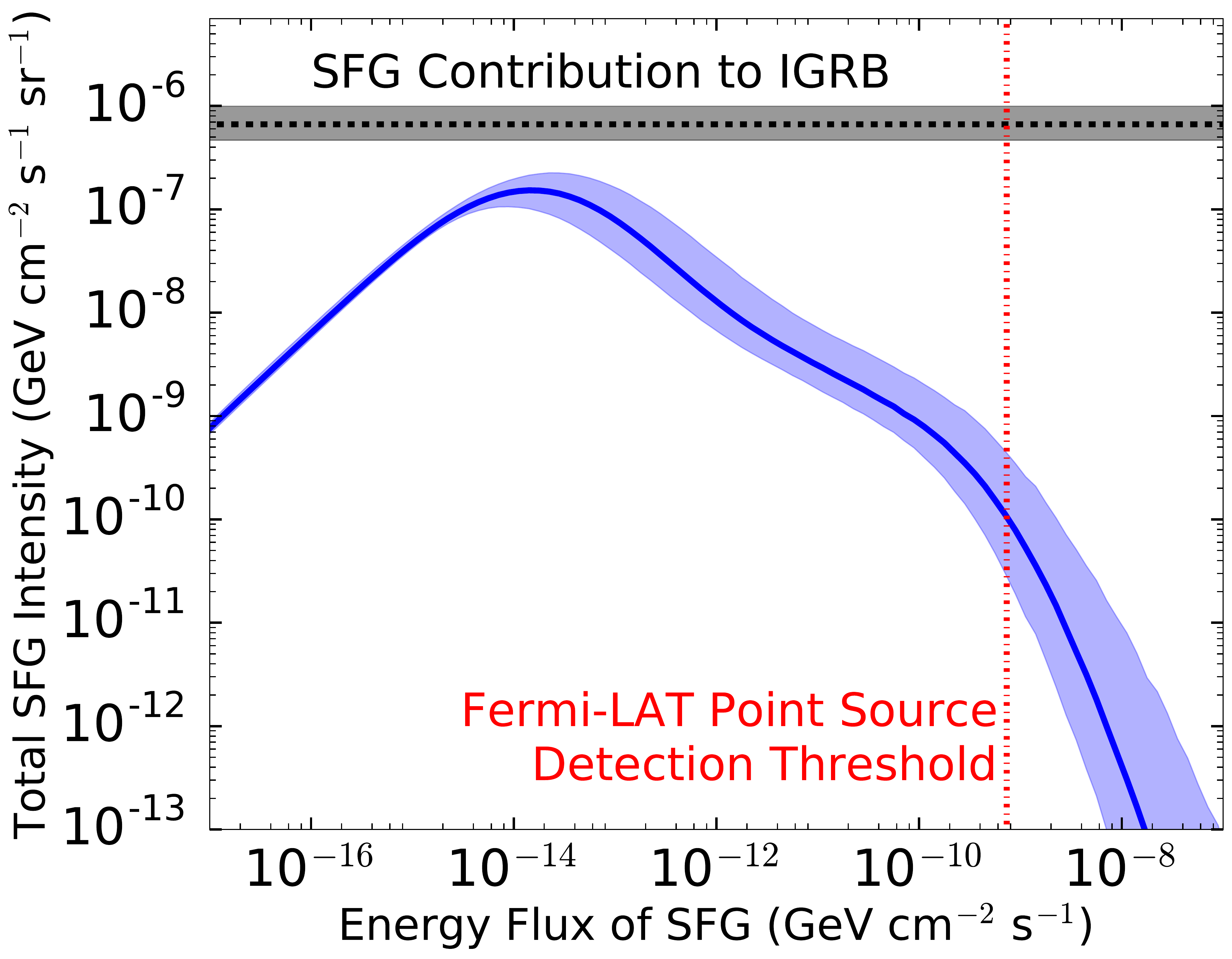}
\caption{The total intensity provided by SFGs as a function of the individual energy flux from each SFG. The total contribution of SFGs to the IGRB is dominated by approximately 3~$\times$~10$^{6}$ systems providing individual energy fluxes on the order of 3~$\times$~10$^{-14}$~GeV~cm$^{-2}$~s$^{-1}$ above an energy of 1~GeV. This is approximately four orders of magnitude below the point-source detection threshold of the Fermi-LAT. Our model also correctly reproduces the expectation that there are a handful of systems above the current Fermi-LAT detection threshold. In this figure (contrary to the rest of the paper), we do not remove the contribution from SFGs which individually exceed the Fermi-LAT point-source detection threshold.} 
\label{fig:fluxdistribution}
\end{figure}

In Figure~\ref{fig:igrbplot} we show the total contribution of SFGs to the IGRB. We find that SFGs contributes 61.0$^{+30.2}_{-18.3}\%$ of the total $\gamma$-ray emission above 1~GeV, and can produce the entirety of the IGRB at energies below 10~GeV. At higher energies the best-fit SFG contribution falls below the IGRB intensity, even when systematic errors in the Galactic foreground contribution are taken into account. However, we note that the contribution of SFGs at high energy depends sensitively on the choice of the spectral index assumed in our models. From an investigation of the four brightest SFGs (NGC 253, NGC 1068, NGC 3034, and NGC 4945) we chose a best-fit spectrum of $\gamma$~=~-2.32 for all SFGs (with a calculated error of 0.06, which is not taken into account in Figure~\ref{fig:igrbplot}). However, there is little evidence that this value is standard for all SFGs, and energy dependent changes in the SFG spectral index would not be well constrained by our algorithm -- which is statistically dominated by $\sim$1~GeV photons.

We note two choices in our modeling which may affect these conclusions. First, we have set a calorimetric limit on the ratio of the $\gamma$-ray luminosity in the 1 --- 100~GeV energy range to be 1.28~$\times$~10$^{-4}$ of the FIR luminosity, following the calculation of~\citep{Lacki:2010vs}. However, we have adjusted this value from the best-fit constraint of~\citep{Lacki:2010vs} (3.1~$\times$~10$^{-4}$), which was calibrated for the energy range 0.1---100~GeV. We made this adjustment using the best fit power-law spectrum of our model (which was not calibrated based on data below 1~GeV). Alterations of this model are possible, and setting a calorimetric limit at 3.1~$\times$~10$^{-4}$ in the 1---100~GeV range increases the best-fit contribution of SFGs to the IGRB to be 64.4$^{+51.7}_{-21.4}\%$. Notably, the lower-limit remains almost unchanged, and the best-fit value increases by only $\sim$10\%. However, the upper-limit increases considerably, since our brightest models have a larger maximum luminosity for systems with large FIR luminosities. 

Second, we assume that the dispersion in the FIR to $\gamma$-ray correlation follows a log-normal distribution that continues to arbitrarily large dispersions. Of our four detected SFGs, we note that none exceeds the median predicted luminosity by more than $\sim$1.5$\sigma$. This is reasonable for a population of four systems. On the other hand, all of these systems are systematically brighter than expected in our best-fit FIR to $\gamma$-ray correlation. This may mean that they are the systems that most significantly exceed the FIR to $\gamma$-ray correlation, and that more extreme deviations from the correlation are suppressed compared to the log-normal expectation. To test the impact of this, we repeat our analysis, but do not allow any system to depart from the FIR to $\gamma$-ray correlation by more than 2$\sigma$. Because this eliminates the brightest SFGs, it decreases the total contribution of SFGs to the IGRB significantly. In this model, SFGs produce only 40.1$^{+13.1}_{-11.2}\%$ of the total IGRB intensity above 1~GeV. 

In Figures~\ref{fig:lgamma_binned_contribution}~and~\ref{fig:fluxdistribution} we show the relative contribution of SFGs to the total IGRB intensity as a function of their FIR luminosity and individual $\gamma$-ray fluxes, respectively. These indicate the characteristics of the SFG populations that are most important to determining the total IGRB intensity from SFGs. In Figure~\ref{fig:lgamma_binned_contribution} we note that the majority of the IGRB is produced by SFGs with FIR luminosities between 10$^{10.5}$~---~10$^{12.5}$~L$_\odot$, consistent with~\citep{Murase:2016gly}. Fortunately, these FIR luminosities are similar to the majority of systems analyzed to produce the FIR to $\gamma$-ray correlation. This indicates that we do not have to extrapolate significantly to smaller or larger SFGs to determine the total contribution of SFGs to the IGRB. In Figure~\ref{fig:fluxdistribution} we show that the majority of the IGRB contribution stems from SFGs with individual fluxes on the order of 10$^{-14}$~GeV~cm$^{-2}$~s$^{-1}$. These systems individually lie nearly four orders of magnitude below the Fermi-LAT detection threshold. This indicates that we do not have to worry about fluctuations in the SFG contribution to the IGRB due to statistical dispersion in a small number of high-flux sources. We additionally find that our model predicts that only a handful of SFGs should be resolved at present --- compatible with observations.

\section{Discussion and Conclusions}

We have produced a novel statistical analysis technique allowing for the combined likelihood analysis of hundreds of $\gamma$-ray point sources in the presence of considerable non-Poissonian background fluctuations. Using this technique, we have calculated the correlation between the FIR and $\gamma$-ray luminosities of 584 SFGs. These techniques go beyond previous efforts~\citep{2012ApJ...755..164A, Rojas-Bravo:2016val} in two ways:  (1) the relationship is not dependent on the (possibly systematically biased) $\gamma$-ray fluxes observed from the LMC and SMC, and (2) the full likelihood profile of low-significance SFGs is examined, allowing for stronger limits on the total SFG emission. Our analysis provides strong evidence for significant dispersion in the FIR to $\gamma$-ray correlation. Despite these additions, we find that the resulting contribution of SFGs to the IGRB is consistent with previous works~\citep{2012ApJ...755..164A, 2014JCAP...09..043T}. Specifically, we find the SFG contribution to the intensity of the IGRB to be 61.0$^{+30.2}_{-18.3}\%$. 

\begin{figure}[tbp]
\centering
\includegraphics[width=.47\textwidth]{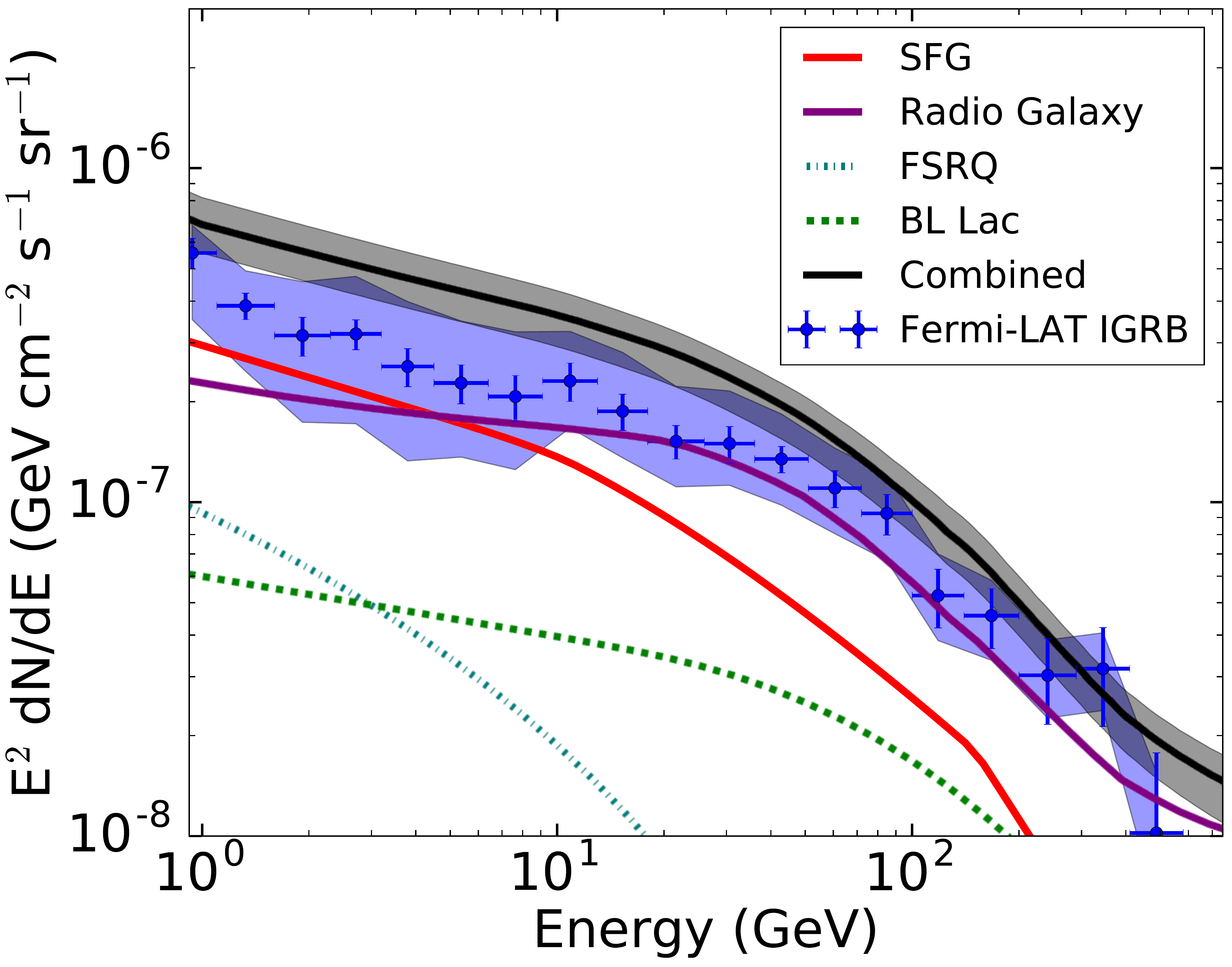}
\caption{The combined contribution of SFGs (as analyzed in this paper), Radio Galaxies (as analyzed in~\citep{Hooper:2016gjy}), and flat spectrum radio quasars and BL Lac objects (as analyzed in \citep{Cholis:2013ena, 2010ApJ...720..435A, 2012ApJ...751..108A, Cuoco:2012yf, Ajello:2013lka}) to the IGRB as observed by the Fermi-LAT. The best-fit spectrum and intensity of each source class is shown, along with the cumulative emission from all source classes. Error bars include the total error on the SFG and Radio galaxy populations, but not uncertainties on the blazar contributions. However, the small flux of blazars makes these uncertainties insignificant. SFGs produce the majority of the total $\gamma$-ray emission below an energy of $\sim$7~GeV, while Radio Galaxies produce the majority of the high energy emission. The total emission from Radio Galaxies and SFGs overproduces the IGRB. However, this lies within the 2$\sigma$ errors from both the subtraction of the Galactic foreground, as well as the 2$\sigma$ uncertainties in the statistical modeling of the SFG and Radio Galaxy contributions.} 
\label{fig:igrbplotradio}
\end{figure}

\subsection{The Composition of the IGRB}
\label{subsec:igrb}
Using a similar analysis technique, we have previously argued that radio galaxies dominate the total intensity of the IGRB, finding a best-fit contribution of 77.2$^{+25.4}_{-9.4}\%$  to the total IGRB intensity above 1~GeV~\citep{Hooper:2016gjy}. This appears to stand in mild tension with the results calculated here. The combination of these contributions places us in the regime of maximum confusion --- SFGs, radio galaxies, and blazars all appear to contribute significantly to the IGRB intensity. The combination of equally important SFG and radio galaxies components appears to be consistent with photon count statistics, which indicate that approximately half of the measured IGRB stems from sources that contribute multiple photons per source~\citep{Zechlin:2015wdz}.

In Figure~\ref{fig:igrbplotradio} we show the combined emission (along with statistical uncertainties) from SFGs (in this paper) and Radio Galaxies (as determined in~\citep{Hooper:2016gjy}). We have also added best-fit estimates for the contribution of BL Lac objects and flat spectrum radio quasars using the analyses of~\citep{Cholis:2013ena, 2010ApJ...720..435A, 2012ApJ...751..108A, Cuoco:2012yf, Ajello:2013lka}, though we do not take the uncertainties in the models into account in the error bars on the total contribution. Figure~\ref{fig:igrbplotradio} demonstrates that these combined contributions overproduce the total IGRB. The tension is somewhat significant at low-energies (due to the combined $\gamma$-ray emission from SFGs and radio galaxies) as well as at high-energies (due to the high luminosity of the radio galaxy model). The tension at high-energies may worsen, as recent analyses of the extragalactic $\gamma$-ray background indicate that blazars produce the majority of this emission~\citep{TheFermi-LAT:2015ykq}. On the other hand, the radio galaxy and SFG contribution at high-energies is very dependent on uncertainties in the spectral modeling of each source, and thus the error bars are likely underestimated.

This tension could be remedied in a number of ways. First, either the FIR to $\gamma$-ray correlation utilized here, or alternatively the radio to $\gamma$-ray correlation utilized in \citep{Hooper:2016gjy} may be offset from the true correlation at the $\sim$2$\sigma$ level. We note that the uncertainty in the SFG contribution is significantly larger, due to the smaller number of systems observed at high statistical significance. 

Second, the correlations may fail for particularly dim or bright SFGs or radio galaxies --- either due to differences in the emission mechanisms of low-luminosity systems, or due to a redshift dependence which alters the evolution of  systems at high redshift. Third, these correlations may fail near the outliers of the correlation --- if we do not allow deviations from the best-fit FIR to $\gamma$-ray correlation for SFGs to exceed 2$\sigma$, the total SFG contribution drops to 40.1$^{+13.1}_{-11.2}\%$. Similarly, as calculated in~\citep{Hooper:2016gjy}, if we disallow deviations exceeding 2$\sigma$ in the radio to $\gamma$-ray correlation for radio galaxies, the best-fit contribution to the IGRB falls to 64.5\%. 

Fourth, these models may \emph{both} be correct in scenarios where the Galactic foreground contribution to the IGRB has been systematically overestimated. The Fermi-LAT collaboration estimates these errors to be roughly 30\% (as shown in the blue shaded error band). This would still be insufficient to completely reconcile our observations of the IGRB with our models of SFGs, radio galaxies and blazars, although it would alleviate much of the tension. It is worth noting that a significant increase in the IGRB intensity from the best-fit valued is disfavored by analyses of the photon statistics near the Fermi-LAT detection limit~\citep{Zechlin:2015wdz}.

A fifth possibility is the degree to which the FIR to $\gamma$-ray correlation (for SFGs) and the radio to $\gamma$-ray correlation (for radio galaxies) double counts the $\gamma$-ray luminosity of the same systems. All radio galaxies certainly have some star-formation activity that contributes to the total $\gamma$-ray flux observed from these galaxies. Additionally, two of the four brightest SFGs (NGC 1068 and NGC 4945) are Seyfert 2 galaxies, and thus may produce significant $\gamma$-ray emission through AGN-like processes. Analyses of the variability and spectrum of NGC 1068 and NGC 4945 indicate that the majority of the $\gamma$-ray emission in NGC 1068 may be produced by the active nucleus, while the $\gamma$-ray emission of NGC 4945 is likely dominated by star formation~\citep{Lenain:2010kc}. However, the potential for bright emission from the population of radio-quiet Seyfert galaxies has also been constrained by~\citep{2012ApJ...747..104A}. The production of significant $\gamma$-ray emission via star formation in ``standard" radio galaxies appears less likely, because a large number of radio galaxies are detected at luminosities far exceeding that expected from star formation activity. However, the proper way to remedy this uncertainty is to attempt a joint-likelihood analysis where the $\gamma$-ray luminosity from each source is set to be the sum of the contribution from the FIR to $\gamma$-ray correlation and the radio to $\gamma$-ray correlation. This will be completed in a forthcoming study. 

\subsection{Implications for the Origin of the IceCube Neutrinos}
\label{subsec:neutrinos}

The culmination of research over the last few years indicates that multiple source classes contribute non-negligibly to the IGRB. This places us in a regime where the source classes responsible for the ultra-high energy cosmic-rays (UHECRs) and the IceCube neutrinos are maximally uncertain. The correlation between the GeV $\gamma$-ray intensity and the implied PeV neutrino and EeV cosmic-ray intensities have focused on three theoretically-motivated arguments: (1) energetics --- which $\gamma$-ray source classes appear to produce a total luminosity sufficient to power the observed neutrino and UHECR fluxes, (2) spectra --- which $\gamma$-ray source classes appear to produce a considerable fraction of their non-thermal emission at the highest energies, (3) angular correlations --- which $\gamma$-ray source classes appear coincident with the observed neutrino and UHECR events.

This work, along with recent papers determining the EGRB (which includes the brightest resolved $\gamma$-ray point sources) contribution from blazars and radio galaxies~\citep{2012ApJ...751..108A, Massaro:2015gla, Lisanti:2016jub, Hooper:2016gjy}, has answered the question of energetics --- all three primary extragalactic source classes (blazars, radio galaxies and SFGs) produce sufficient $\gamma$-ray emission to power the UHECR and neutrino fluxes.  

The measurements of the $\gamma$-ray source spectra, on the other hand, are significantly less certain for two reasons. The first is the large extrapolation between ``Fermi-LAT energies" ($\sim$0.1---1000~GeV), the IceCube energy scale ($\sim$100~TeV) and the UHECR energy scale ($\sim$10$^{9}$~GeV). It is reasonable that some (or all) of these emission sources may have a cutoff energy, over which their emission becomes exponentially suppressed. The second, and less-appreciated, complication is the systematic issues in determining the Fermi-LAT $\gamma$-ray spectrum at high energies. Because the photon flux from all sources drops precipitously at high energies, the spectral fit to a Fermi-LAT point source is almost always dominated by the lowest $\gamma$-ray energies under consideration. Thus, it is difficult to determine the spectral shape of SFGs in our analysis above an energy of $\sim$10~GeV. 

However, more detailed analyses have utilized GeV $\gamma$-ray spectral information to constrain the contribution of IGRB sources to the IceCube diffuse neutrino flux~\citep{Tamborra:2014xia, Murase:2015xka, Bechtol:2015uqb}. A study by \citep{Bechtol:2015uqb} argues that SFGs can produce no more than 28\% of the IceCube neutrino flux, when the SFG spectrum is marginalized over current $\gamma$-ray uncertainties. However, this analysis is sensitive to systematic errors in that spectral determination. For a \mbox{$\gamma$~=~-2.3} spectrum, SFGs can produce no more than 10\% of the IceCube events, while for a spectrum \mbox{$\gamma$~=~-2.2,} this rises to 30\% of the IceCube flux at 100~TeV, and over 60\% of the neutrino flux at 1~PeV. A recent analysis by ~\citep{Hooper:2016jls} found that the spectrum of radio galaxies is consistent (at the 1-2$\sigma$ level) with the spectrum needed to explain the diffuse IceCube neutrino flux.

Finally, angular correlations can be utilized to constrain the contribution of a specific $\gamma$-ray source class with high-energy radiation, utilizing the assumption that the brightest sources at $\gamma$-ray energies (and in multiwavelength catalogs) are also among the brightest neutrino sources. This has been used to constrain the neutrino flux from blazars to less than 20\% of the total diffuse neutrino flux~\citep{Glusenkamp:2015jca, Murase:2016gly}. Making similar correlations in the time domain has constrained the total contribution from $\gamma$-ray bursts to provide less than 1\% of the observed neutrino emission~\citep{2012Natur.484..351A}. The strong constraints on blazars and $\gamma$-ray bursts put significant pressure on the two most likely photohadronic neutrino sources, where the source contribution to the IGRB can be suppressed by the hard $\gamma$-ray spectrum. 

The difficulty in utilizing hard spectrum photohadronic sources to produce the IceCube neutrino flux makes the precise determination of the IGRB spectrum more pertinent. If blazars and $\gamma$-ray bursts are not capable of explaining the neutrino flux -- then SFGs and radio galaxies (which dominate the IGRB, but produce a small fraction of the remaining EGRB) become prime candidates to explain the IceCube emission. In this context, the nearly equivilent contribution of SFGs and radio galaxies to the IGRB makes the IceCube neutrinos even more disturbing. The intensity of the IceCube neutrino flux is approximately 7$\times$10$^{-8}$~GeV~cm$^{-2}$~s$^{-1}$ at an energy of 1~PeV, compared to a IGRB flux of 2$\times$10$^{-7}$~GeV~cm$^{-2}$~s$^{-1}$ at an energy of 10~GeV. Taking into account the roughly 3/2 ratio of the neutrino flux compared to the $\gamma$-ray flux for purely hadronic processes, this implies that the IGRB sources must have an average $\gamma$-ray spectrum of E$^{-2.13}$ in order to explain the neutrino emission~\citep{Murase:2013rfa}. We note that necessary spectrum would be even harder if a portion of the $\gamma$-ray emission stems from non-hadronic processes such as inverse-Compton scattering~\citep{Chakraborty:2012sh}. This spectrum is harder than the best-fit $\gamma$-ray spectrum found for both radio galaxies~\citep{Hooper:2016gjy} and SFGs.

If both of these source classes individually produce only 50\% of the IGRB, the situation becomes more dire. If a source produces only $\sim$50\% of the IGRB intensity, but the entirety of the neutrino flux, the necessary $\gamma$-ray spectrum hardens to nearly E$^{-2}$. In \citep{Murase:2013rfa}, this spectral argument is employed to argue that a source class that dominates the IceCube neutrino flux must produce at least 30-40\% of the IGRB background. Alternatively, multiple source classes could contribute to the diffuse neutrino flux in similar ratios to their IGRB contribution. However, this would require some hardening (compared to the best-fit models) of both radio galaxies and SFGs. This may occur in scenarios where cascades inside of the sources themselves provide a significant portion of the $\gamma$-ray flux, although TeV detections of M82 and NGC 253, by VERITAS and H.E.SS, respectively, appear to argue against significant hardening in the SFG spectrum~\citep{Acciari:2009wq, Acero:2009nb, 2011ApJ...734..107L}. 

As a final option, it is possible that the sources of IceCube neutrinos may reside in extremely dense environments, that are opaque to $\gamma$-ray emission ~\citep{Tamborra:2015fzv, Maggi:2016bbi, Murase:2015xka, Senno:2015tsn}. In this scenario, the large IceCube neutrino flux can be accomodated without significant contributions to the IGRB. More investigations are needed (for example of the auto-correlation of IceCube neutrino events~\citep{Silvestri:2009xb, Murase:2016gly} in order to constrain this possibility. 

The tension described above could be almost directly transferred to a discussion of UHECRs~\citep[see e.g. ][]{Murase:2014tsa, Alvarez:2016otl}. Given the relative maturity of IGRB observations (compared to those of neutrinos and UHECRs), it is clear that continued $\gamma$-ray observations are necessary to shed light on this riddle. In particular, the source-to-source dispersion in the $\gamma$-ray spectra of radio galaxies, SFGs and blazars may allow for an enhanced emission component at high-energies. It is finally worth noting the possibly important contribution of TeV instruments such as H.E.S.S, VERITAS, and HAWC to understanding the diffuse neutrino flux.  

\subsection{Final Conclusions}
Interestingly, we note that in the models we have produced, SFGs produce the majority of the $\gamma$-ray emission below $\sim$7~GeV, while radio galaxies produce the majority of $\gamma$-ray emission above this energy. Intriguingly, this energy range is similar to the dip in the cross-correlation of the anisotropy energy spectrum, as measured in~\citep{Fornasa:2016ohl}. It is difficult to straightforwardly assign this dip to populations of SFGs, as their anisotropy is expected to lie far below the sensitivity of the Fermi-LAT (see Figure~\ref{fig:fluxdistribution}). Alternatively, the rapidly declining flat-spectrum radio quasar contribution is capable of producing considerable anisotropy at low-energies, while remaining subdominant in its contribution to the total IGRB intensity. Further studies are necessary to combine the anisotropic and intensity measurements of the IGRB in order to produce a model which explains all available data. 

Finally, we note that the statistical methods produced here can be utilized to detect the combined emission from any $\gamma$-ray source population that has a theoretically motivated prior for the $\gamma$-ray flux from each individual source, e.g. from dark matter annihilation in the population of the Milky Way dwarf spheroidal galaxies. Additionally, this technique could be expanded into regimes where the diffuse background is likely to vary significantly (e.g. regions near the Galactic plane, or near the Fermi bubbles), as long as an estimate for the background flux, now denoted as P$_{i, bg}$, can be calculated for each $\gamma$-ray point source residing in region $i$. Future works will explore the extension of this computational method to an improved understanding of $\gamma$-ray emitting sources.

\appendix

\section{A Null Test of the (-$\ell$, -b) Sky Positions}
\label{sec:nulltest}

\begin{figure}[tbp]
\centering
\includegraphics[width=.48\textwidth]{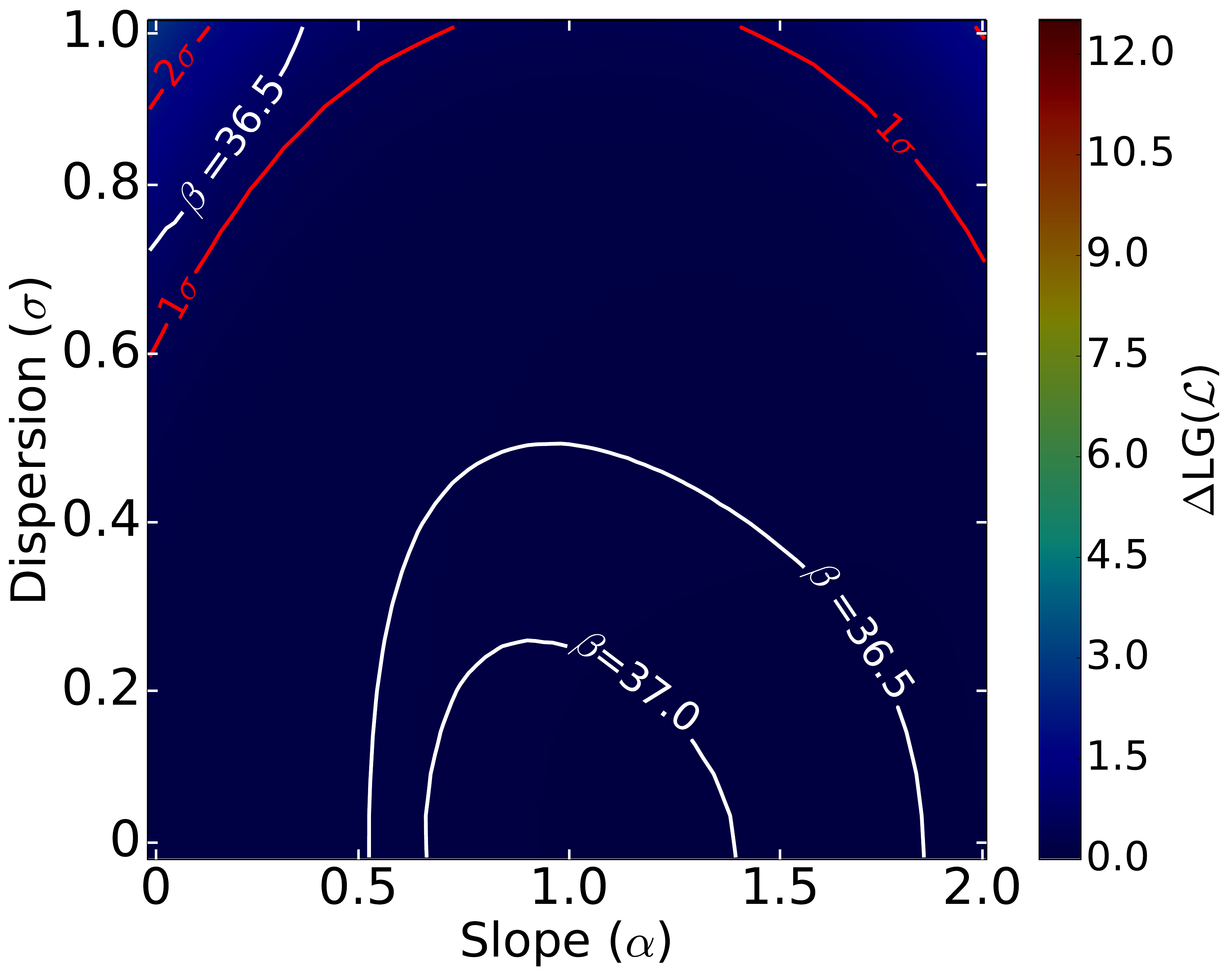}
\caption{The best fit values of $\alpha$, $\beta$ and $\sigma$ for an analysis where the SFGs are placed at the negative sky locations (-$\ell$, -b), compared to the true location of each SFG. As expected, no emission is detected and the best-fit is only $\Delta$LG($\mathcal{L}$)=0.08 compared to a model with no $\gamma$-ray flux corresponding to fake SFGs. This demonstrates that the detected dispersion from SFGs in our default analysis is due to the characteristics of the SFGs themselves and not background fluctuations. The dispersion in the $\gamma$-ray flux due to background systematics is properly taken into account by P$_{bg}$ in our model.}
\label{fig:nulltest}
\end{figure}

Thus far, we have not utilized the blank sky positions (-$\ell$, -b) in our model of the diffuse background fluctuations (P$_{bg}$). Now we will utilize this set of 584 sky positions to test the analysis methods described in Section~\ref{sec:models}. Here, we analyze the Fermi-LAT data for each SFG, pretending the SFG resides at its negative sky position. We then run the resulting best-fit $\gamma$-ray fluxes through our pipeline to determine the best-fit values for $\alpha$, $\beta$ and $\sigma$. In Figure~\ref{fig:nulltest} we show, as expected, that there is no statistical preference for $\gamma$-ray emission from these blank-sky locations, finding a best-fit $\Delta$LG($\mathcal{L}$ of only 0.08, compared to a model with no $\gamma$-ray flux. This finding was relatively obvious, as their is no $\gamma$-ray emission emanating from these point sources.

More importantly, there is no evidence for dispersion in the FIR to $\gamma$-ray correlation in these sky positions. This result is somewhat less-obvious, because we know that there are variations in the $\gamma$-ray flux at each sky position resulting from background fluctuations in the Fermi-LAT data. Our analysis technique is designed to account for these fluctuations by comparing them with the flux distribution of background fluctuations (P$_{bg}$). Note that these (-$\ell$, -b) are not includedd in our calculation of P$_{bg}$, and thus their background fluctuations are independent. Thus, this test provides evidence that our analysis technique is working properly. This, in turn, provides evidence that the observed dispersion in the FIR to $\gamma$-ray correlation is due to an intrinsic flux dispersion in the SFGs themselves, rather than an unaccounted for variation in the $\gamma$-ray flux from background mismodeling.

\section{Reconstructing an Injected $\gamma$-ray Signal}
\label{sec:injectedsignal}

\begin{figure*}[tbp]
\centering
\includegraphics[width=.48\textwidth]{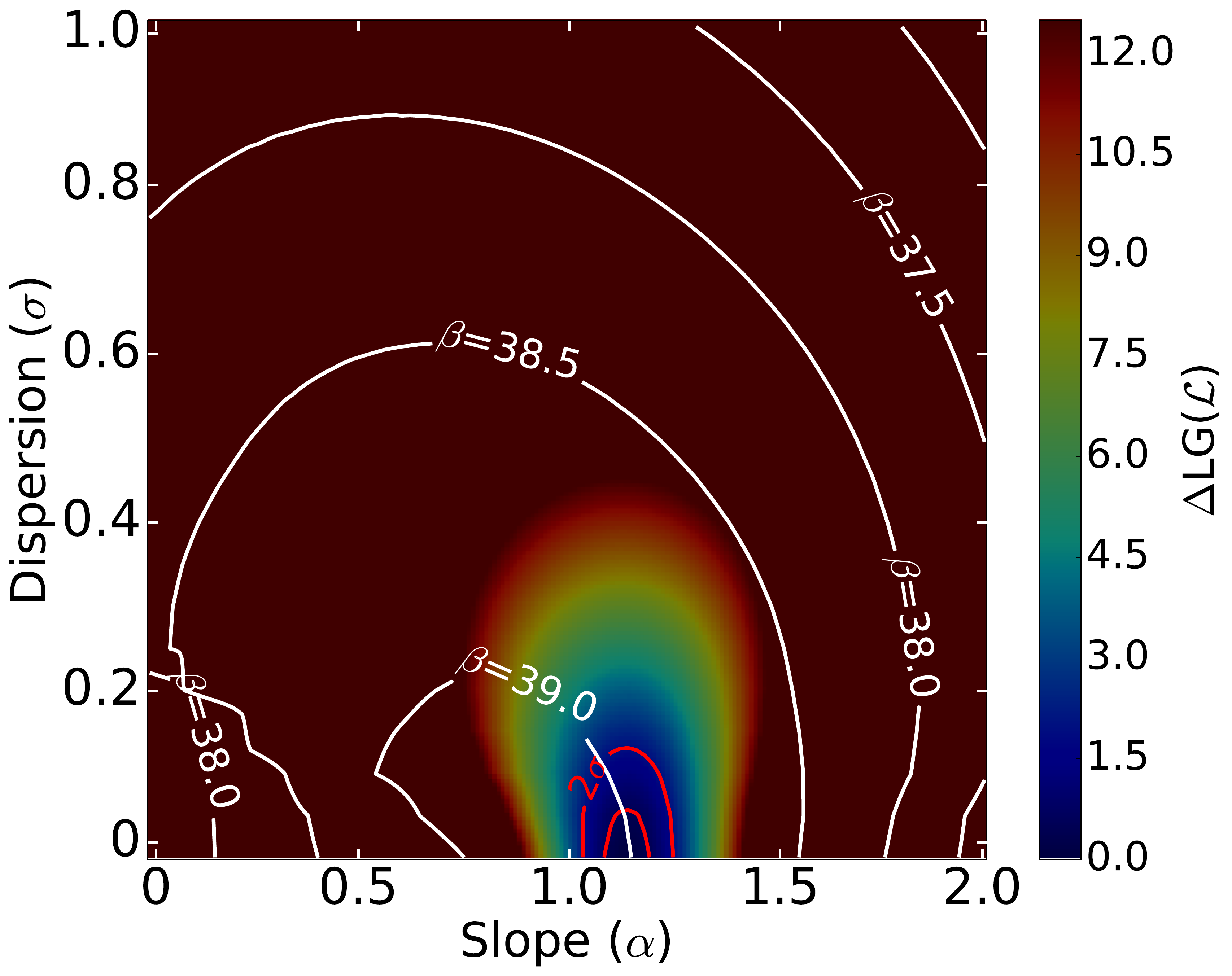}
\includegraphics[width=.48\textwidth]{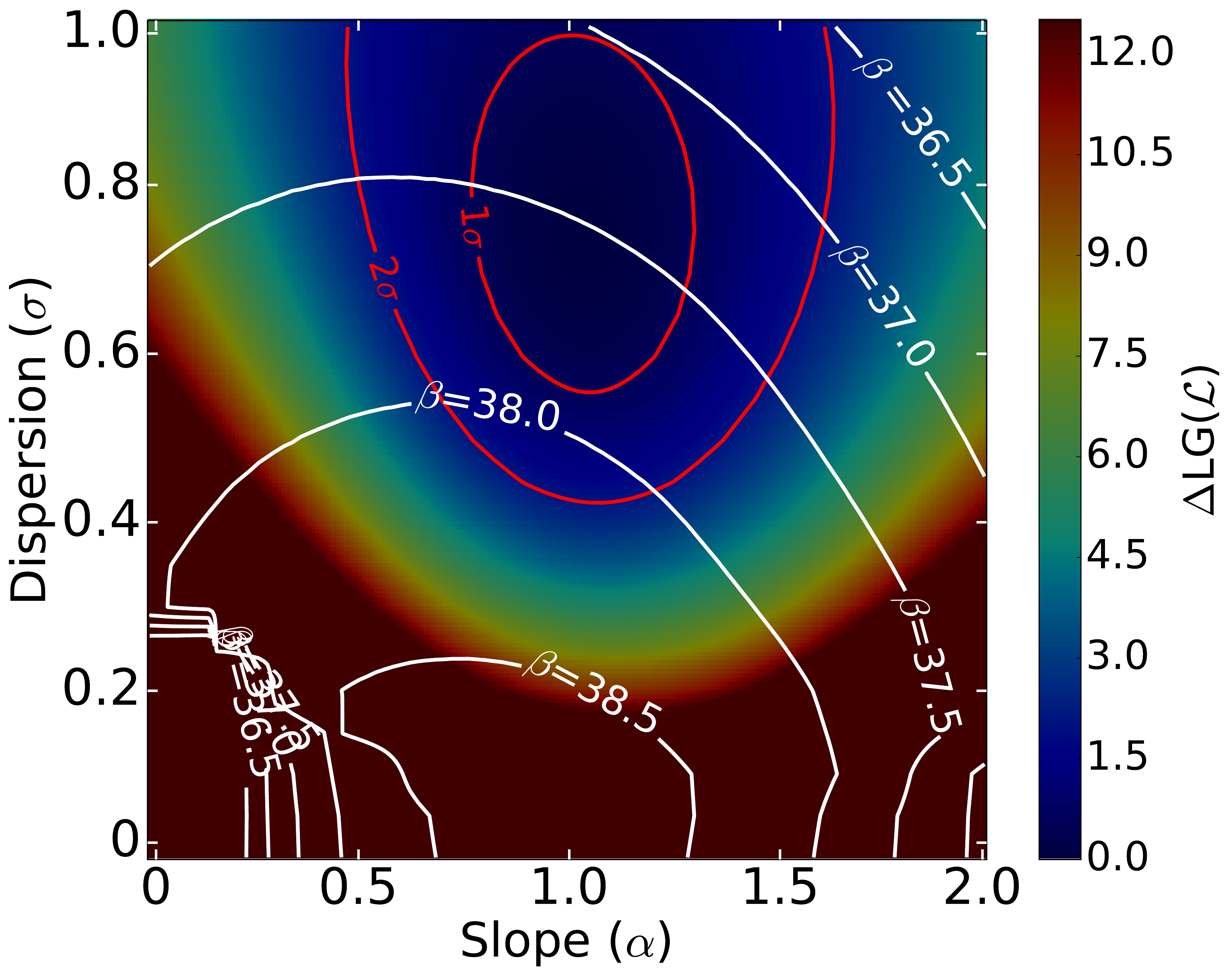}

\caption{A reconstruction of the SFG FIR to $\gamma$-ray correlation produced by analyzing the blank sky positions (-$\ell$, -b) after a $\gamma$-ray signal has been injected following the correlation log$_{10}$(L$_\gamma$ / (erg s$^{-1}$)~=~1.17~log$_{10}$(L$_{IR}$/ 10$^{10}$L$_\odot$) + 38.985 with no dispersion. The best-fit values of $\alpha$, $\beta$ and $\sigma$ are reconstructed via our pipeline for our default analysis (left) and for an analysis where negative background fluxes were not considered (right). The default analysis correctly reproduces the FIR to $\gamma$-ray correlation. It is noteworthy that no intrinsic dispersion is found in the reconstructed analysis, indicating that the fluctuations are being properly accounted for via the P$_{bg}$ distribution. On the contrary, considering only positive fluctuations in the background fluxes fails to reconstruct the injected relationship, significantly decreasing the value of $\beta$ and increasing the dispersion in the FIR to $\gamma$-ray correlation. Further details in the text.}
\label{fig:injection}
\end{figure*}

In addition to utilizing the (-$\ell$, -b) sky positions in a null-test for the FIR to $\gamma$-ray correlation, we can test the analysis by injecting a fake SFG signal at the (-$\ell$, -b) sky positions and analyzing th resulting signal with our pipeline. Specifically we utilize the {\tt gtobssim} tool to produce a simulated photon flux at all 584 negative SFG positions in our default model. The spectrum of this injected signal is given by -2.32, and the intensity is correlated to the observed FIR luminosity of each positive SFG via the relationship log$_{10}$(L$_\gamma$)~=~1.17log$_{10}$(L$_{IR}$) + 38.985, with L$_\gamma$ and L$_{IR}$ being normalized to 1 erg s$^{-1}$ and 10$^{10}$~L$_\odot$ respectively. In this test, we assume that there is no source-to-source dispersion in the FIR to $\gamma$-ray relation. We then add the photons produced by {\tt gtobssim} to the real Fermi-LAT data surrounding the sky positions (-$\ell$, -b). We then utilize our point-source algorithm (within the Fermi tools) to calculate the best-fit $\gamma$-ray flux at the sky positions (-$\ell$, -b). Note again that the (-$\ell$, -b) sky-positions have not been utilized to produce the P$_{bg}$ background model in our analysis, and thus the $\gamma$-ray background fluctuations at these sky positions are independent of our P$_{bg}$ model.

In Figure~\ref{fig:injection} (left) we plot the output of our analysis, and demonstrate that we successfully extract the injected signal. Marginalizing over all variables, we recover best-fit values of $\alpha$~=~1.11~$\pm$~0.05, $\beta$~=~38.99~$\pm$~0.05, $\sigma$~=~0.06~$\pm$~0.05, a result that is within $\sim$1.5$\sigma$ in its reconstruction of $\alpha$, and almost exactly recovers the value of $\beta$. We note that since $\sigma$ must be non-negative, the mean value of sigma in our reconstruction is likely to be pulled away from 0. Indeed we find the best-fitting value of $\sigma$ resides at $\sigma$~=~0.0. The fact that our best-fit reconstruction is consistent with no dispersion in the L$_\gamma$ to L$_{IR}$ correlation is significant, as it demonstrates that the flux dispersion due to background fluctuations is being properly removed by the analysis. This test thus provides further evidence that the dispersion observed in the SFG population stems from differences in the $\gamma$-ray luminosity of the SFGs themselves. 

\section{The Importance of Negative Background Fluctuations}

\begin{figure*}[tbp]
\centering
\includegraphics[width=.48\textwidth]{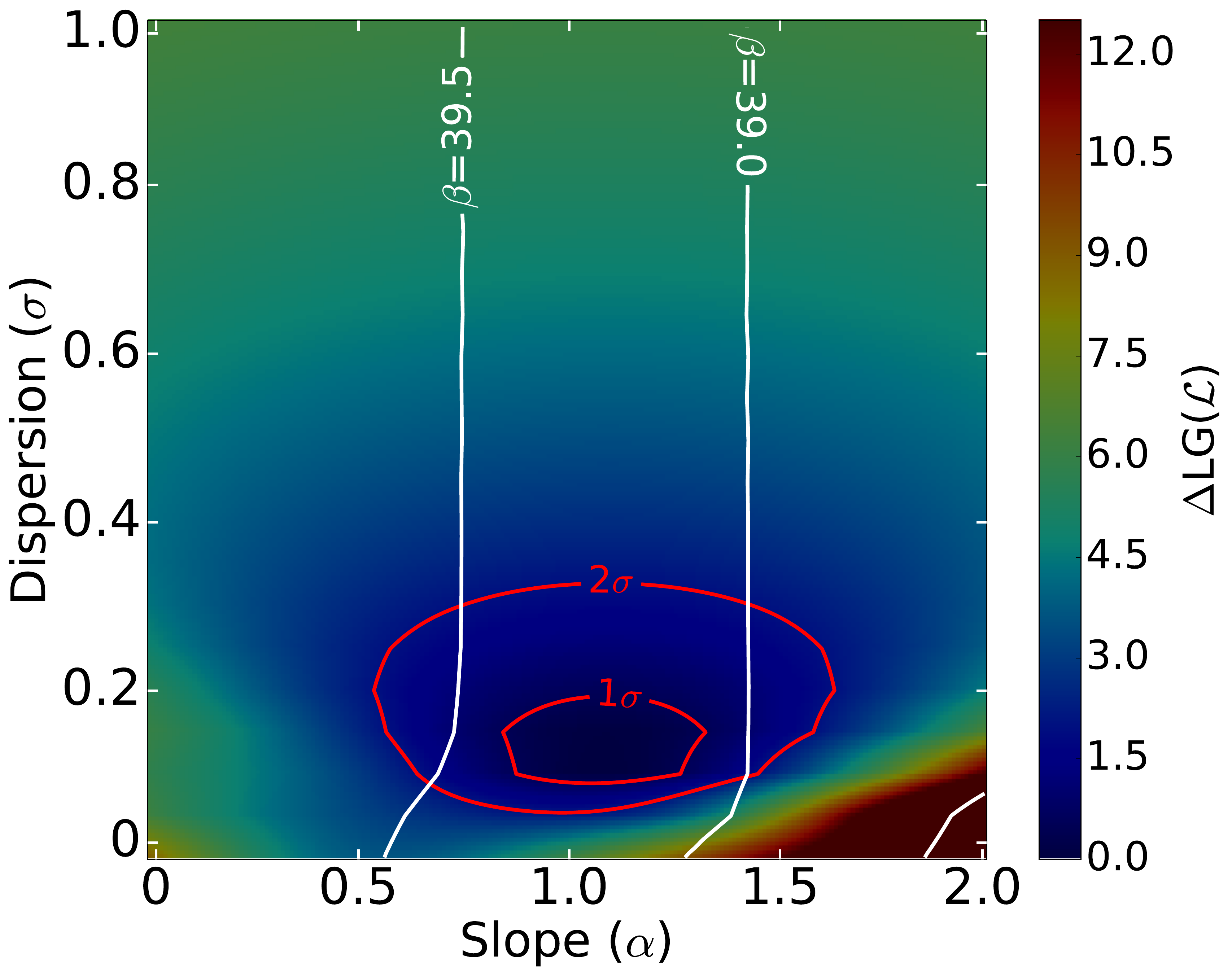}
\includegraphics[width=.48\textwidth]{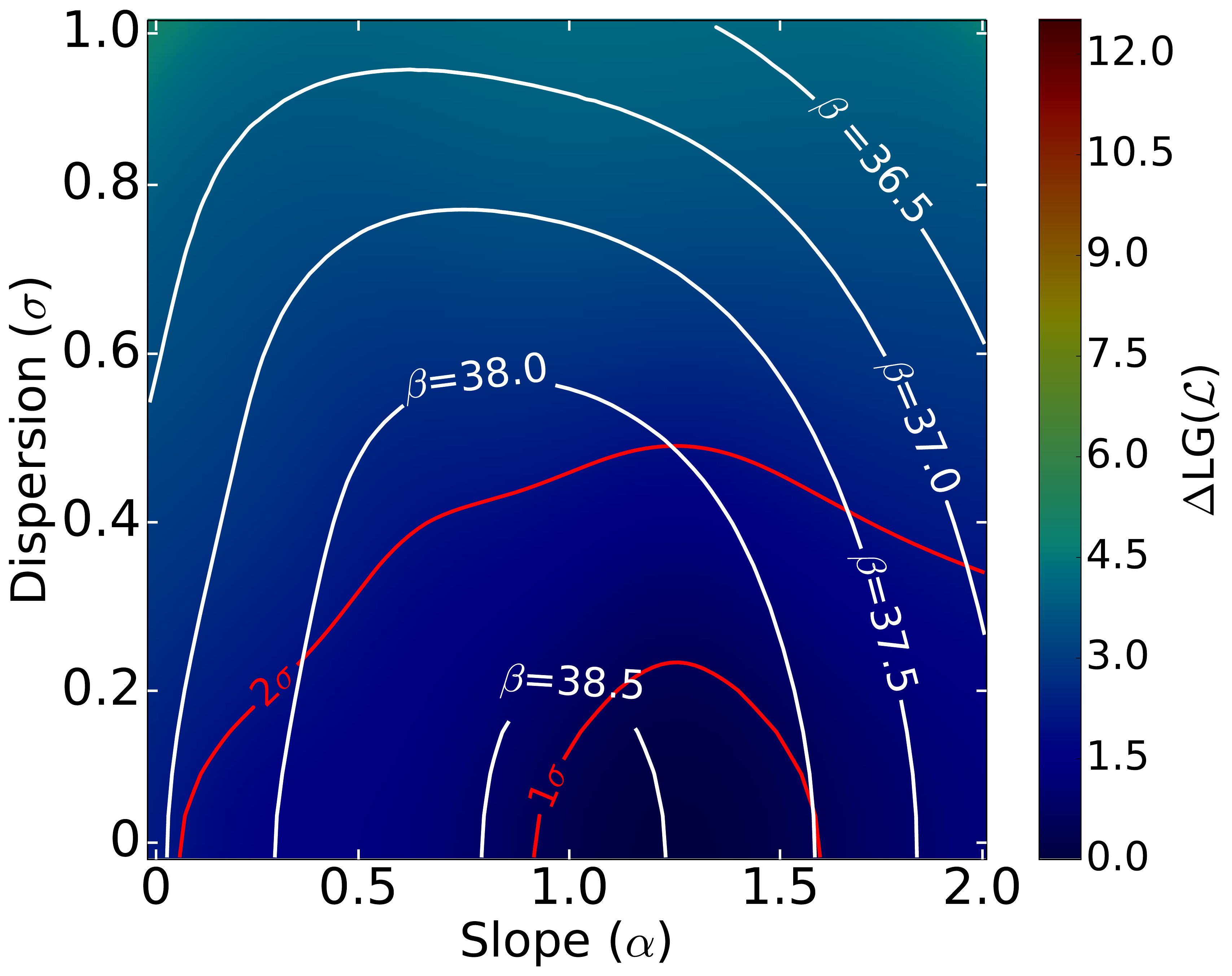}

\caption{A reconstruction of the SFG FIR to $\gamma$-ray correlation produced by analyzing only the four systems (NGC 253, NGC 1068, NGC 3034, NGC 4945) that have highly significant detected $\gamma$-ray fluxes (left), and the ensemble of 580 systems that do not have individually statistically significant detections (right). It is apparent that a significant portion of the statistical correlation comes from the brightest systems. However, the brightest systems do not accurately reproduce the calculated from the full-ensemble of SFGs, indicating the importance in correctly modeling the low-significance systems. The IR to $\gamma$-ray correlation is still detected in the ensemble of non-detected SFGs, at a total statistical significance of $\sim$3$\sigma$.}
\label{fig:detected}
\end{figure*}

\begin{figure*}[tbp]
\centering
\includegraphics[width=.48\textwidth]{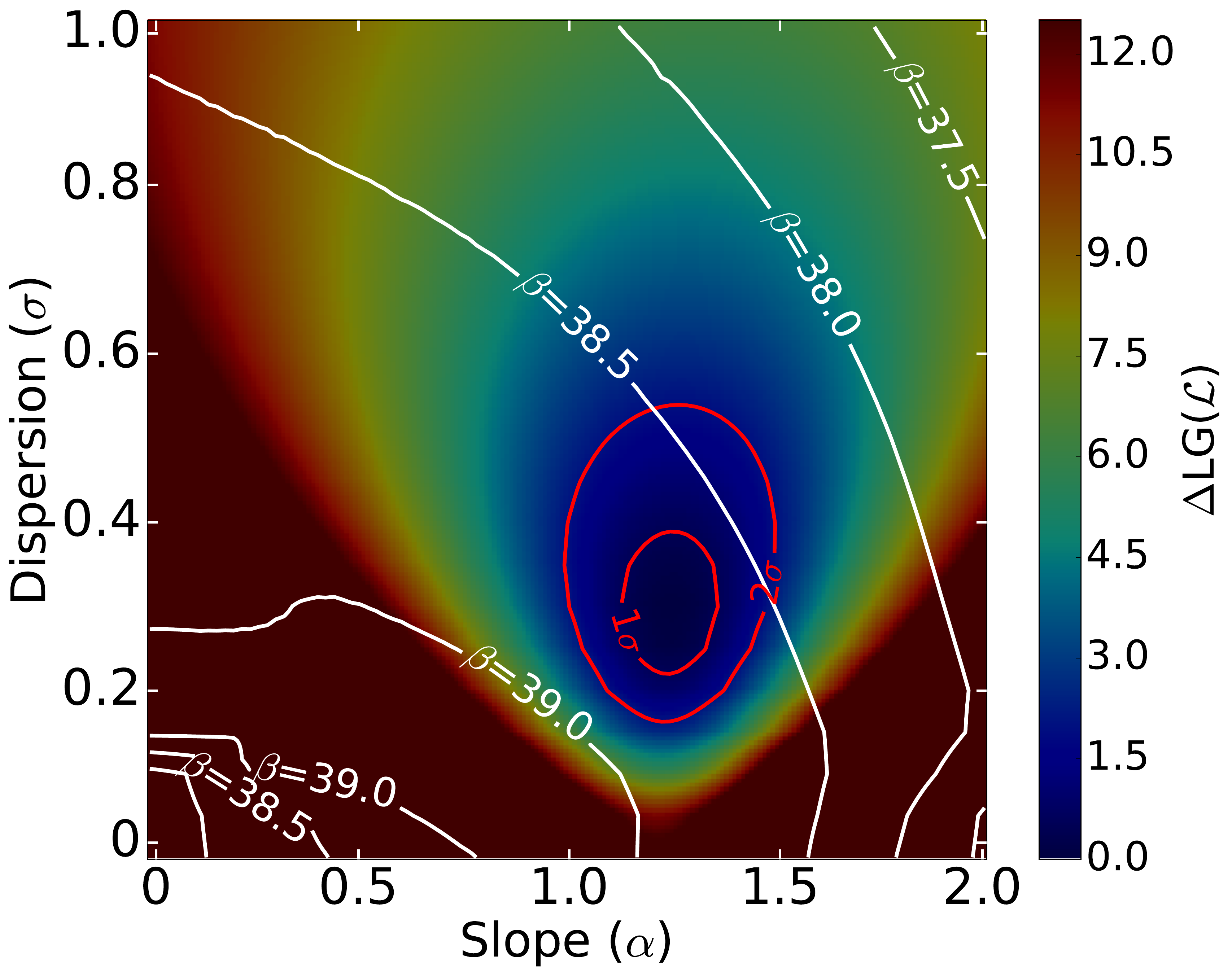}
\includegraphics[width=.48\textwidth]{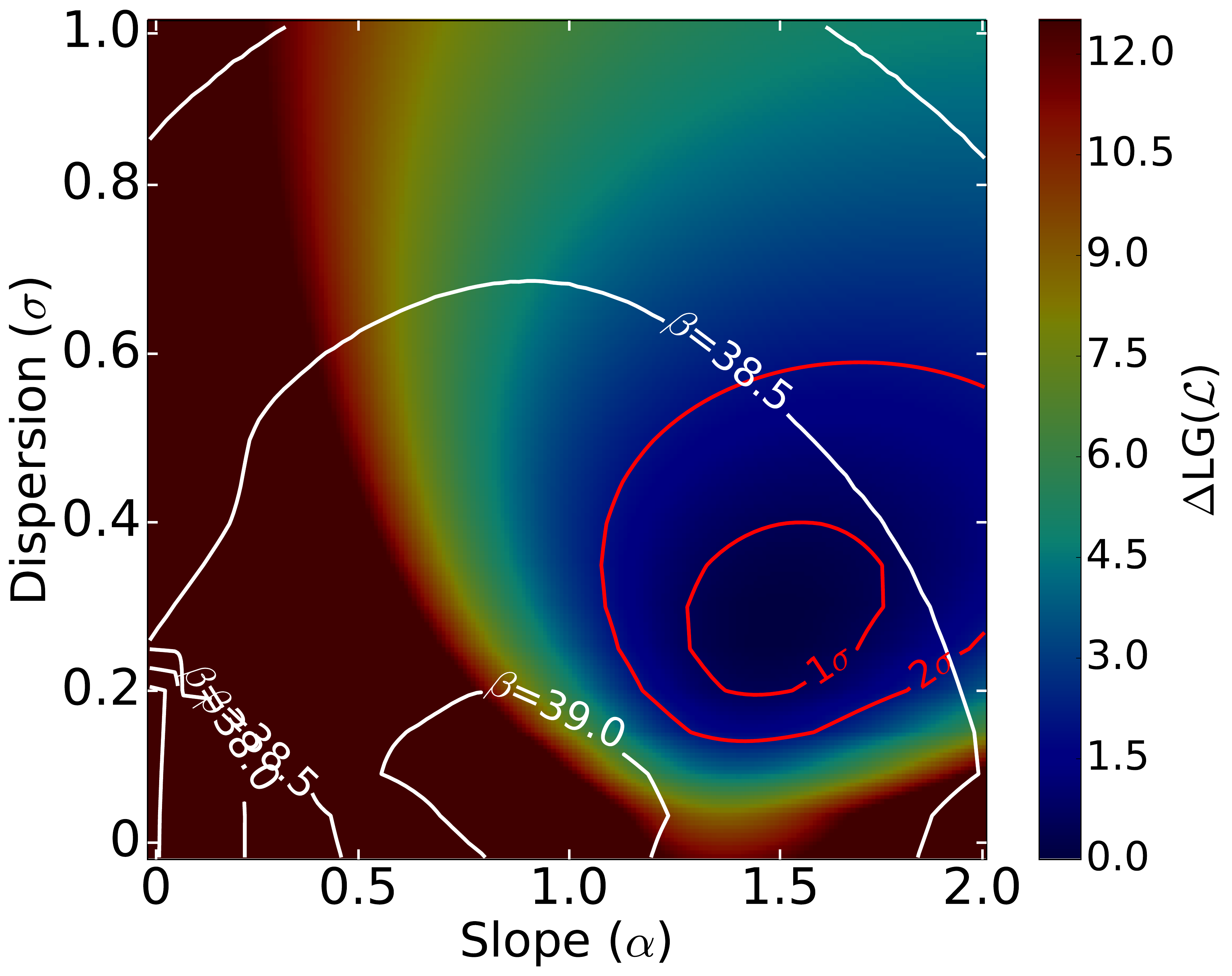}

\caption{A reconstruction of the SFG FIR to $\gamma$-ray relationship produced by analyzing only the 50 systems that are expected to provide the brightest $\gamma$-ray signals. (Left) The 50 brightest systems are correctly chosen using the value $\alpha$~=~1.17 which was detected by the full analysis. (Right) The 50 brightest systems are chosen incorrectly assuming that $\alpha$~=~0.5, which biases our analysis to the nearest SFGs. While the left reconstruction provides a fairly accurate (though systematically brighter) representation of the full SFG population, an incorrect choice in the SFG population does not provide sufficient information to model the IR to $\gamma$-ray correlation. Thus, the observation of 50 systems is inadequate without underlying knowledge of the FIR to $\gamma$-ray correlation.}
\label{fig:brightest50}
\end{figure*}

In typical joint-likelihood analyses of the Fermi-LAT data, only point-sources with positive flux are considered as components of the non-Poissonian $\gamma$-ray background fluctuations. This is, of course, theoretically well-motivated as physical point-sources can only contribute non-negative $\gamma$-ray fluxes. Moreover, many of the brightest unaccounted for fluctuations reside from point-source that lie just below the Fermi-LAT detection threshold, and thus produce only positive fluctuations~\citep{Carlson:2014nra}.

However, this technique induces biases when analyzing the flux contribution of very dim $\gamma$-ray point sources. The reason is straightforward --- if a true $\gamma$-ray point source is coincident with a positive fluctuation in the $\gamma$-ray background then the treatment of non-Poissonian background fluctuations can be taken into account. The significance of the point source can be decreased, based on the probability that some, or all, of the $\gamma$-ray flux results from the background contribution. In Fermi-LAT analyses, this is typically done in TS space, where the statistical significance of point sources are re-fit based on the distribution of TS values found when fitting blank sky positions.  

However, a true $\gamma$-ray point source may also lie coincident with a negative fluctuation of the $\gamma$-ray background. As seen in Figure~\ref{fig:Pbg}, this background fluctuation may even exceed the physical point-source flux. If the blank sky technique only evaluates non-negative $\gamma$-ray background fluctuations, it will attribute the observed $\gamma$-ray point source to be coincident with a null fluctuation in the $\gamma$-ray background, attributing a systematically smaller flux to the true $\gamma$-ray point source. When a joint-likelihood analysis is employed on the resulting data, the constraint will be systematically strengthened by the unphysically low upper limit derived for the point source lying coincident with a significant negative fluctuation of the gamma-ray background.

In Figure~\ref{fig:injection} (right) we follow the same analysis routine described in Section~\ref{sec:models}, but produce a model for P$_{bg}$ that considers only non-negative point-source fluxes at the null sky positions utilized in our analysis. In this case we find that our analysis grossly mismodels the resulting FIR to $\gamma$-ray relationship. In particular, the best-fitting value of $\beta$ decreases by more than an order of magnitude, while the value of $\sigma$ significantly increases to produce a log-normal dispersion of nearly an order of magnitude. 

The reasons for this are clear. The lack of negative background fluctuations produces multiple systems which have very strong upper limits on the $\gamma$-ray flux, because the positive point-source is coincident with a significantly negative background fluctuation. This drives the mean ($\beta$) of the correlation to an anomalously low value. To account for this, the bright point sources (stacked on top of positive background fluctuations) must be significantly dispersed from their best-fit values, driving the value of $\sigma$ up. The importance of properly dealing with negative background fluctuations in the Fermi-LAT data has important consequences for many analyses, e.g. the joint-likelihood analysis of dwarf spheroidal galaxies. 

\section{The Contribution from sub-threshold SFGs}
\label{sec:appendixsubthreshold}
It is worth questioning (in particular at this moment) whether the analysis of hundreds of SFGs in this analysis provides useful additional information, or whether the FIR to $\gamma$-ray correlation is described entirely by the few brightest systems in our analysis. In Figure~\ref{fig:detected} (left) we plot the best-fit values of $\alpha$, $\beta$ and $\sigma$ for the four SFGs that are detected at high statistical significance in our sample (NGC 253, NGC 1068, NGC 3034, NGC 4945). We find that this small sample produces a correlation that is significantly biased towards bright $\gamma$-ray SFGs, with an average best-fit $\beta$~$\approx$39.3. This exceeds the best-fit value determined by our full analysis by more than half an order of magnitude, even after accounting for the lower value of $\sigma$. Additionally, the analysis of only bright systems does not produce a particularly strong constraint on $\alpha$, compared to the full analysis. In Figure~\ref{fig:detected} (right), we find that the stacked population of non-detected SFGs obtains a smaller (but non-zero) correlation between the FIR and $\gamma$-ray luminosities, which is statistically significant at the 3$\sigma$ level. In particular we find a best fit value $\alpha$~$\approx$~1.25, $\beta$~$\approx$~38.6 and $\sigma$~$\approx$~0.0. While the statistical significance of the FIR to $\gamma$-ray correlation is smaller in non-detected systems compared to the four brightest systems, we note that the best-fit values of the correlation are nearly as close to the best-fit values in our full analysis. 

The true importance of the additional 580 SFGs in our analysis is seen in the marked difference between Figure~\ref{fig:detected} (left) and our default result in Figure~\ref{fig:sfgplot} (left). In particular the existence of systems that are not $\gamma$-ray bright significantly lowers the best-fit value of the normalization, $\beta$, increases the dispersion, $\sigma$, and slightly alters the slope of the correlation, $\alpha$. We note that if the likelihood profile of non-detected systems are not taken into account the best-fit value for the dispersion ($\sigma \approx 0.15$) is consistent with the dispersion identified by \citep{2012ApJ...755..164A}. Interestingly the dispersion is also small among non-detected systems (though it is not strongly constrained). The increase in dispersion appears to stem from the offset between the best-fit values of the FIR to $\gamma$-ray correlation observed in detected systems, versus the smaller values found in non-detected systems. 

In Figure~\ref{fig:brightest50} we instead analyze the contribution from a sub-population of 50 SFGs that are likely to produce the brightest $\gamma$-ray signal. This provides us a relatively diverse population that includes all four of the brightest SFGs, as well as a reasonable population of non-detected SFGs that (due to their bright predicted $\gamma$-ray flux) can significantly constrain the FIR to $\gamma$-ray relation. Of course, a priori, it is difficult to determine which 50 SFGs will be brightest, without knowledge of the FIR to $\gamma$-ray correlation. We first (left) select the correct sub-sample of the 50 systems with the highest predicted $\gamma$-ray flux, using the value $\alpha$~=~1.17 that is preferred by our full analysis. In this case we find that the brightest 50 SFGs provide most of the relevant information --- although the value of $\beta$ in this analysis is still somewhat higher than found in our default analysis. In FIgure~\ref{fig:brightest50} (right), we instead use an incorrect subset of the 50 brightest SFGs, using a predicted relation $\alpha$~=~0.5 to select the brightest systems. We note that this provides an inaccurate calculation of the FIR to $\gamma$-ray correlation. The system is not biased towards the value of $\alpha$ we selected. Instead, this subset of systems does not contain sufficient information to constrain the model. Thus, even a sample of 50 systems would be insufficient to model the FIR to $\gamma$-ray correlation, without additional knowledge of the expected result.

\section{Interpreting the Significance of Point Sources in the Context of P$_{bg}$}
\label{sec:statsignificance}

Finally, it is worthwhile to comment on the use of P$_{bg}$ to determine the statistical probability that a given $\gamma$-ray flux actually stems from the physical object targeted in the search at a given sky position. The background fluctuations which make up the P$_{bg}$ spectrum are produced by the sum of all sub-threshold emission sources in the $\gamma$-ray sky, including not only diffuse background mismodeling, but also sub-threshold populations of blazars, pulsars, radio galaxies, star-forming galaxies, supernovae remnants, and possibly other unknown sources.  Thus, by calculating the probability that a given $\gamma$-ray flux is produced by a background fluctuation P$_{bg}$, we are also calculating the inverse probability -- that the $\gamma$-ray point source is not caused by any typical fluctuation in the $\gamma$-ray background. If the chosen sky position is not thought to be unique in any other way besides the existence of the SFG in question, then this acts as the probability that a given source is producing a $\gamma$-ray signal in the presence of background fluctuations. This statistical analysis, of course, fails in the case that multiple unique sources are thought to be coincident on the $\gamma$-ray sky. In this case, the sky position is not well represented by P$_{bg}$, and P$_{bg}$ cannot provide us with additional information. For example, this method would be unable to provide information about which of the two coincident sources produces the observed $\gamma$-ray flux.

\section*{Acknowledgements}
We thank Andrea Albert, John Beacom, Keith Bechtol, Mauricio Bustamante, Alex Drlica-Wagner, Mattia Fornasa, Dan Hooper, Adam Leroy, Shirley Li, Kohta Murase, Irene Tamborra, and Todd Thompson for many helpful conversations that improved the quality of this paper, and acknowledge support from NSF Grant PHY-1404311 to John Beacom.

\bibliography{sfgs}

\end{document}